\documentclass[twoside]{IEEEtran}

\newcommand{\sqr}{\vrule height6pt width6pt depth1pt}
\def\qed{\hfill\sqr}

\usepackage{graphics} 
\usepackage{epsfig} 
\usepackage{times} 
\usepackage{amsmath} 
\usepackage{amssymb}  
\usepackage{setspace}
\usepackage{color}

\newtheorem{theorem}{Theorem}
\newtheorem{lemma}{Lemma}

\definecolor{dred}{rgb}{0,0,0}
\def\tdred{\textcolor{dred}}

\begin{document}

\title{ \huge On the Deterministic Code Capacity Region
of an Arbitrarily Varying Multiple-Access Channel Under List Decoding}


\author{Sirin~Nitinawarat,~\IEEEmembership{Member,~IEEE}
\thanks{S. Nitinawarat is with the Department of Electrical
and Computer Engineering at the University of Illinois at
Urbana-Champaign, IL, USA.  The work of S. Nitinawarat was
conducted partly at the University of Maryland, while it was
supported by the National Science Foundation under Grants
CCF0515124, CCF0635271, CCF0830697. The material in this paper was
presented in part at the IEEE International Symposium on
Information Theory, Austin, Texas, USA, June 13-18, 2010.
} }

\maketitle

\begin{abstract}
We study the capacity region $C_L$ of an arbitrarily varying
multiple-access channel (AVMAC) for deterministic codes with
decoding into a list of a fixed size $L$ and for the average error
probability criterion. Motivated by known results in the study of
fixed size list decoding for a point-to-point arbitrarily varying
channel, we define for every AVMAC whose capacity region for
random codes has a nonempty interior, a nonnegative integer $\Omega$
called its symmetrizability. It is shown that for every $L \leq
\Omega$, $C_L$ has an empty interior, and for every $L \geq (\Omega+1)^2$,
$C_L$ equals the nondegenerate capacity region of the AVMAC for
random codes with a known single-letter characterization.  For a
binary AVMAC with a nondegenerate random code capacity region, it
is shown that the symmetrizability is always finite.
\end{abstract}


\section{Introduction}

We study the deterministic code capacity region of an arbitrarily
varying multiple-access channel (AVMAC) under fixed size-$L$ list
decoding.
For every received sequence, a list decoder outputs a list of
message pairs of size at most $L$.  The error occurs when the
transmitted message pair is not in the output list. We restrict
ourselves to a discrete memoryless AVMAC with finite inputs,
output and state alphabets and the average error probability
criterion.

For a point-to-point transmission over an arbitrarily varying
channel (AVC), it is known \cite{Ahlswede_78} that the
(list-of-$1$ size) deterministic code capacity equals either 0 or
the random code capacity.  The latter capacity is defined for a
``random code'' in which the encoder and the decoder are assumed
to have a shared access to a random experiment of which the result
can be used in selecting a deterministic code from a pool of them.
A sufficient condition was introduced in \cite{Ericson_85} for the
deterministic code capacity to be zero; this condition, of the AVC
being ``symmetrizable,'' was shown to be necessary as well for the
AVC to have a zero deterministic code capacity
\cite{Csiszar_Narayan_88}.
When list decoding of a fixed size $L$ is considered, it also
holds that the list-of-$L$ size capacity for deterministic codes
equals either 0 or the random code capacity; a necessary and
sufficient condition for the list-of-$L$ size capacity for
deterministic codes to be zero was given in
\cite{Blinovsky-Narayan-Pinsker_95, Hughes_97} in terms of a
quantity called the ``symmetrizability'' of the AVC defined in
\cite{Hughes_97}.  This concept of the ``symmetrizability'' of an
AVC can be regarded as a generalization of the condition of the
AVC being symmetrizable considered in \cite{Ericson_85,
Csiszar_Narayan_88}.  Precisely, an AVC is symmetrizable if its
symmetrizability is at least 1.

Next, turning to transmission over an AVMAC, with the usual
decoding ($L$ = 1), Jahn \cite{Jahn_81} showed that the capacity
region for deterministic codes $C_1$ either has an empty interior
or equals the random code capacity region defined and
characterized therein. Gubner \cite{Gubner_90} introduced the
condition of the AVMAC being symmetrizable and showed that it
implies that the interior of $C_1$ is empty.  Later, Ahlswede and
Cai \cite{Ahlswede-Cai_99} proved that this condition is also
necessary for the emptiness of the interior of $C_1$.

In the present paper, we introduce a concept of symmetrizability
of an AVMAC and study its relationship with its list-of-$L$ size
capacity region for deterministic codes

\section{Preliminaries}

We start with the definitions of the discrete memoryless AVMAC and
describe certain quantities relating to its specific behavior.

Let $\mathcal{X}, \mathcal{Y}, \mathcal{Z}$ and $\mathcal{S}$ be
finite sets representing the two input alphabets, the output
alphabet and the state alphabet, respectively.  The AVMAC is
determined by a family of conditional probability distributions
$W(z|x, y, s)$ on $\mathcal{Z} (z \in \mathcal{Z})$, defined by
two input signals $x \in \mathcal{X},\ y \in \mathcal{Y}$ and a
state $s \in \mathcal{S}$. It is assumed that the AVMAC is
memoryless, i.e., that the transition probability function
$W^n({\bf z} | {\bf x}, {\bf y}, {\bf s}),\ {\bf x} = (x_1,
\ldots, x_n) \in \mathcal{X}^n,\ {\bf y} = (y_1, \ldots, y_n) \in
\mathcal{Y}^n,\ (z_1, \ldots, z_n) \in \mathcal{Z}^n,\ {\bf s} =
(s_1, \ldots, s_n) \in \mathcal{S}^n$ satisfies $W^n({\bf z} |
{\bf x}, {\bf y}, {\bf s}) = \prod_{t = 1}^n W(z_t | x_t, y_t,
s_t)$.  We denote such a channel as $T=(W, \mathcal{X},
\mathcal{Y}, \mathcal{S}, \mathcal{Z})$.  A deterministic code
$(\mathcal{U}^{(n)}, \mathcal{V}^{(n)})$ of length $n$ and
cardinalities $M_1$, $M_2$ is a collection of $\mathcal{U}^{(n)} =
\{{\bf x}_1, \ldots, {\bf x}_{M_1}\} \subseteq {\mathcal{X}^n} $
and $\mathcal{V}^{(n)} = \{{\bf y}_1, \ldots, {\bf y}_{M_2}\}
\subseteq {\mathcal{Y}^n}$. We call $R_1 = \frac{1}{n}
\log_2{M_1}$ and $R_2 = \frac{1}{n} \log_2{M_2}$ the {\em rates}
of the codes for transmitter 1 and transmitter 2, respectively, and
$(R_1, R_2)$ the {\em rate-tuple} of the code.

In this paper, we consider list decoding of a fixed size $L$; the
usual decoding corresponds to the special case of $L = 1$. Given
$M_1$ and $M_2$, a list-of-$L$ size decoder $\phi_L$ is defined,
for every ${\bf z} \in \mathcal{Z}^n,$ as $\phi_L \left( {\bf z }
\right) \subseteq 
\tdred{\{1, \ldots, M_1\} \times \{1, \ldots, M_2\}}$ such that 
$\vert \phi_L \left( {\bf z} \right) \vert \leq L.$ 
The code together with the list decoder $\mathcal{C}_L =
(\tdred{\mathcal{U}^{(n)}, \mathcal{V}^{(n)}}, \phi_L)$ is called a deterministic code
decoded into a list of size $L$. The error probability of decoding
into a list of size $L$ when the message pair $(i,j)$ is
transmitted over the AVMAC in the state ${\bf s} \in
\mathcal{S}^n$ is defined as
\begin{eqnarray}
    e_L(i, j, {\bf s}) &=& e_L(i, j, {\bf s}, \mathcal{C}_L) \nonumber \\
    &\triangleq& \sum_{{\bf z} \in \mathcal{Z}^n:\, (i,j) \notin \phi_L({\bf z})}
        W^n({\bf z}| {\bf x}_i, {\bf y}_j, {\bf s}),
\end{eqnarray}
and the corresponding average error probability is defined as
\begin{equation}
    \bar{e}_L({\bf s})\ =\ \bar{e}_L({\bf s}, \mathcal{C}_L)\ \triangleq\
    \sum_{i = 1}^{M_1} \sum_{j = 1}^{M_2} \frac{1}{M_1 M_2}
    e_L(i, j, {\bf s}).
\end{equation}
For $R_1 >0, R_2 >0$, we are interested in the quantity
\[
\hspace{-2.7in}
\bar{p}_L(R_1, R_2) \]
\begin{equation}
\nonumber
    =\  \sup\limits_{\epsilon > 0}~
    \limsup\limits_{n \rightarrow \infty}~
    \mathop{\min_{\mathcal{C}_L,\ \log_2{M_1} \geq (R_1 - \epsilon) n}}
    _{\ \ \ \ \log_2{M_2} \geq (R_2 - \epsilon)n}
    ~\max_{{\bf s} \in \mathcal{S}^n}~ \bar{e}_L({\bf s}, \mathcal{C}_L).
\end{equation}
Define the {\em list-of-$L$ size capacity region} $C_L = C_L(T)$
of $T$ for deterministic codes under the average error probability
criterion to be the closure of the region $\{(R_1 \geq 0, R_2 \geq
0):\ \bar{p}_L(R_1, R_2) = 0\}$; let $int(C_L)$ denote the
interior of $C_L$.

The random code capacity region $C^R$, defined in \cite{Jahn_81},
will play a central role in this paper.  $C^R$ was characterized
therein as the closure of the convex hull of the following region
\begin{equation}
    \mathop{\bigcup_{P_X(x),}}_{P_Y(y)}
    \left \{ (R_1, R_2):
    \begin{array}{ll}
        0 \leq R_1 \leq \inf_{P_S(s)}
                 I(X \wedge Z | Y) \\
        0 \leq R_2 \leq \inf_{P_S(s)}
                 I(Y \wedge Z | X) \\
        R_1 + R_2 \leq \inf_{P_S(s)}
                 I(X, Y \wedge Z)
    \end{array} \right \},
    \label{eqn-rand-cap}
\end{equation}
with the union being over all distributions $P_X$ on $\mathcal{X}$
and $P_Y$ on $\mathcal{Y}$ and with the joint distribution of $(X,
Y, Z) \in \mathcal{X} \times \mathcal{Y} \times \mathcal{Z}$ being
$P_{XYZ}(x, y, z) = P_X(x)P_Y(y) \sum_{s \in \mathcal{S}} P_S(s)
W(z|x, y,s)$.

\section{Main Results}

The following theorem extends the statement of Jahn \cite{Jahn_81}
from $L = 1$ to the case $L \geq 1$.

\vspace{0.05in}
\begin{theorem} Either $C_L$ equals $C^R$ or $~int(C_L) =
\emptyset$.
\end{theorem}
\vspace{0.05in}

The proof of Theorem 1 follows the derivation in \cite{Jahn_81}.
When $~int(C_L) \neq \emptyset$, a short deterministic prefix code
with decoding into a list of size $L$, at a rate-tuple with each
individual rate being nonzero, can be concatenated with a
collection (polynomial ensemble size) of long codes to show that
$C_L \supseteq C^R$.  That $C_L \subseteq C^R$ also follows, upon
noting that $C^R$ remains unchanged for list decoding of a fixed
size, in a similar manner to the AVC case
\cite{Blinovsky-Narayan-Pinsker_95, Hughes_97} and an exercise in
\cite[p. 230]{Csiszar_Korner_81}.

\vspace{0.05in} \textbf{Definition 1:}  For a positive integer
$u$, an AVMAC $T$ is {\em u-symmetrizable} if at least one of the
following holds.

a) There exists a conditional probability distribution\\
\tdred{$U$ from $\mathcal{X}^u \times \mathcal{Y}^u$ to $\mathcal{S}$}
such that for any $x_1, \ldots, x_{u+1} \in \mathcal{X},\
y_1, \ldots, y_{u+1} \in \mathcal{Y},\ z \in \mathcal{Z}$ and any permutation $\pi$ on $[u+1] \triangleq
\{1, \ldots, u+1\}$,
\begin{equation}
    \hspace{-0.5in} \sum_{s \in \mathcal{S}} W(z | x_1, y_1, s)
                U(s | x_2, y_2, \ldots, x_{u+1}, y_{u+1}) = \nonumber
\end{equation}
\begin{equation}
    \label{sym-1}
    \sum_{s \in \mathcal{S}} W(z | x_{\pi(1)}, y_{\pi(1)}, s)
                U(s | x_{\pi(2)}, y_{\pi(2)}, \ldots, x_{\pi(u+1)}, y_{\pi(u+1)}).
\end{equation}

b) For some integers $a, b \geq 0$ satisfying $(a+1)(b+1)
\geq u+1$, there exists a conditional probability distribution\\
\tdred{$U$ from $\mathcal{X }^a \times \mathcal{Y}^b$ to $\mathcal{S}$} 
such that for any $x_1, \ldots,
x_{a+1} \in \mathcal{X},\ y_1, \ldots, y_{b+1} \in \mathcal{Y},\ s
\in \mathcal{S},\ z \in \mathcal{Z}$, and any permutations $\pi$
on $[a+1]$ and $\sigma$ on $[b+1]$,
\begin{equation}
    \hspace{-0.4in} \sum_{s \in \mathcal{S}} W(z | x_1, y_1, s)
                U(s | x_2, \ldots, x_{a+1}, y_2, \ldots, y_{b+1}) = \nonumber
\end{equation}
$\sum_{s \in \mathcal{S}} W(z | x_{\pi(1)}, y_{\sigma(1)}, s)$
\begin{eqnarray}
    U(s | x_{\pi(2)}, \ldots, x_{\pi(a+1)}, y_{\sigma(2)}, \ldots, y_{\sigma(b+1)}).
    \label{sym-2}
\end{eqnarray}
To simplify terminology, we take all AVMACs to be 0-symmetrizable.
It is clear that if $T$ is $u$-symmetrizable, then $T$ is also
$u'$-symmetrizable for all $0 \leq u' \leq u$.  The
symmetrizability of $T$ denoted by $\Omega = \Omega(T)$ is defined as the
largest integer $u$ for which $T$ is $u$-symmetrizable.

\tdred{Note that the symmetrizable condition in \cite{Gubner_90} is tantamount to the 
1-symmetrizable condition here.}

\vspace{0.05in}
\begin{theorem} For an AVMAC with symmetrizability $\Omega$,
$~int(C_L) = \emptyset$ for every $L \leq \Omega$.
\end{theorem}
\vspace{0.05in}

\begin{theorem}
For an AVMAC with symmetrizability $\Omega$
and for every $L \geq (\Omega+1)^2$, $C_L$ equals $C^R$ with a
nonempty interior.
\end{theorem}
\vspace{0.05in}

A natural question that arises at this point is whether there
exists an AVMAC with unbounded symmetrizability, i.e., it is
$u$-symmetrizable for an infinitely many values of $u$.  If such
an AVMAC also satisfies $int \left( C^R \right) \neq \emptyset,$
then it follows from Theorem 2 that $C_L \neq C^R,$ for every $L
\geq 1.$  Our last result shows that for a binary AVMAC $\left(
\vert \mathcal{X} \vert = \vert \mathcal{Y} \vert = \vert
\mathcal{S} \vert = \vert \mathcal{Z} \vert = 2 \right),$ this
contingency never arises.  Furthermore, it is shown that the
symmetrizability of a binary AVMAC can be arbitrarily large.

\vspace{0.05in}
\begin{theorem}
Any binary AVMAC satisfying $int \left( C^R \right) \neq \emptyset$ must
have bounded symmetrizability.  Moreover, for any $N > 0,$ there exists a
binary AVMAC with the symmetrizability larger than $N$.
\end{theorem}

\section{\tdred{Proofs}}

For a positive integer $M$, let $[M]$ denote $\{1, \ldots, M\}$, and for a set $K \subseteq [M] \times [M]$, let $I_K \triangleq
\{i \in [M]: \exists j \in [M] \mbox{~such~that}~(i,j) \in K \}$
and, similarly, $J_K \triangleq \{j \in [M]: \exists i \in [M]
\mbox{~such~that}~(i,j) \in K \}$.
\tdred{
We shall call a set $K \subseteq [M] \times [M]$ a {\em diagonal} if 
$|K| = | I_K | = |J_K|$ and call it a {\em rectangle} if $|K| =
|I_K||J_K|$; when the size of $K$ is specified, say $|K| = A$, we shall also refer to them as $A$-diagonal and $A$-rectangle respectively.
}

\vspace{0.05in} \textbf{Proof of Theorem 2:}  For a fixed $L \leq
\Omega$ and any $\delta > 0$, we consider any deterministic code
(decoded into a list of size $L$) $\mathcal{C}_L =
(\mathcal{U}^{(n)}=\{{\bf x}_i\}_{i=1}^M,\
\mathcal{V}^{(n)}=\{{\bf y}_i\}_{i=1}^M,\ \phi_L({\bf z}))$ with
$R=\frac{1}{n}\log_2{M} \geq \delta$.  By Definition 1, either
(\ref{sym-1}) holds with $u = \Omega$ or (\ref{sym-2}) holds with some
$a, b$ such that $(a+1)(b+1) = \Omega+1$, or both.

First, suppose that (\ref{sym-1}) holds with $u = \Omega$. For any 
\tdred{
$\Omega$-diagonal $K
= \{(i_1, j_1), (i_2, j_2), \ldots, (i_{\Omega}, j_{\Omega})\} \subset [M] \times
[M]$, with $(i_1, j_1) < (i_2, j_2) < \ldots < (i_{\Omega}, j_{\Omega})$ (for a
fixed ordering of $[M] \times [M]$),
} 
let $S_K$ denote a random state sequence with
distribution $U^n({\bf s} | {\bf x}_{i_1}, {\bf y}_{j_1}, \ldots,
{\bf x}_{i_{\Omega}}, {\bf y}_{j_{\Omega}} )$. Also, for any $(i, j) \in [M]
\times [M]$, let

\vspace{0.07in} $E[W^n({\bf z} | {\bf x}_i, {\bf y}_j, S_K)]
\triangleq$
\[\sum_{{\bf s} \in \mathcal{S}^n} W^n({\bf z} | {\bf x}_i, {\bf y}_j, {\bf s})
U^n({\bf s} | {\bf x}_{i_1}, {\bf y}_{j_1}, \ldots, {\bf x}_{i_{\Omega}},
{\bf y}_{j_{\Omega}}).\] Then, for any \tdred{${\Omega}+1$-diagonal $K'$ and 
one of its element $(i^*, j^*)$,} 
we have

\vspace{0.07in}
\tdred{
$\sum_{(i,j) \in K'}
        E[e_L(i, j, S_{K' \backslash \{(i, j)\}})]$
\begin{align}
&=\  \sum_{ (i,j) \in K' }
    \left ( 1 -
      \mathop{\sum_{{\bf z}: (i, j)}}_{\ ~~~\in \phi_L({\bf z})}
      E[W^n({\bf z} | {\bf x}_{i}, {\bf y}_{j}, S_{K' \backslash \{(i, j)\}})] \right ) \nonumber \\
&=\  (\Omega+1) - \nonumber \\
& \ \ \ \   \sum_{{\bf z}} 
\mathop{\sum_{(i,j) \in K' :}}_{(i, j) \in
\phi_L({\bf z})}
      E[W^n({\bf z} | {\bf x}_{i}, {\bf y}_{j}, 
      S_{K' \backslash \{(i, j)\}})]      \nonumber \\
&=\  (\Omega+1) - \nonumber \\
& \ \ \ \   \sum_{{\bf z}} 
\mathop{\sum_{(i,j) \in K':}}_{(i, j) \in \phi_L({\bf z})}
      E[W^n({\bf z} | {\bf x}_{i^*}, {\bf y}_{j^*}, 
      S_{K' \backslash \{(i^*, j^*)\}})]      \mbox{~by~}(\ref{sym-1}) \nonumber \\
&\geq\  (\Omega+1) - L,\ \ \ \ \mbox{by~}|\phi_L({\bf z})| \leq L. \label{eqn:Proof-Thm2-1}
\end{align}}
\indent
Next, let 
\tdred{$\mathcal{P}_{\Omega}$ be the set of all $\Omega$-diagonals in $[M] \times [M].$}
Then, $|\mathcal{P}_{\Omega}| = {\tiny \left(
\begin{array}{cc} M \\ {\Omega} \end{array} \right)^2 {\Omega}!}$ and

\vspace{0.1in}
$\frac{1}{|\mathcal{P}_{\Omega}|} \sum_{K \in \mathcal{P}_{\Omega}}
    E[\bar{e}_L(S_K)]$
\begin{eqnarray}
    &\geq&
    \frac{1}{|\mathcal{P}_{\Omega}| M^2} \sum_{K \in \mathcal{P}_{\Omega}}
    \sum_{(i,j) \in I_K^c \times J_K^c} E[e_L(i, j, S_K)] \nonumber \\
    &=&
    \frac{1}{|\mathcal{P}_{\Omega}| M^2} \sum_{K' \in \mathcal{P}_{{\Omega}+1}}
    \sum_{(i,j) \in K'} E[e_L(i, j, S_{K' \backslash \{(i,j)\}})] \nonumber \\
    &\geq&  \frac{|\mathcal{P}_{{\Omega}+1}|({\Omega}+1)}{|\mathcal{P}_{\Omega}| M^2} (1 - \frac{L}{{\Omega}+1}),\
            \mbox{~by~}(\ref{eqn:Proof-Thm2-1})  \nonumber \\
    &=& (\frac{M-{\Omega}}{M})^2(1-\frac{L}{{\Omega}+1}). \nonumber
\end{eqnarray}
Then, for any $M=\lfloor 2^{\delta n} \rfloor$,\\ $\liminf_{n
\rightarrow \infty} \frac{1}{|\mathcal{P}_{\Omega}|} \sum_{K \in
\mathcal{P}_{\Omega}} E[\bar{e}_L(S_K)] > 0$ if $L \leq {\Omega}$. Since the
left side is an average of $\bar{e}_L({\bf s})$ with respect to a
distribution of $S_K$ with $K$ being uniform on $\mathcal{P}_{\Omega},$
we get that $\liminf\limits_{n \rightarrow \infty}
\max\limits_{{\bf s}} \bar{e}_L \left( {\bf s} \right) > 0$ if $L
\leq {\Omega}$. It now follows that $~int(C_L) = \emptyset,$ as $\delta >
0$ can be taken to be arbitrarily small.

Next, consider the case in which (\ref{sym-2}) holds with some $a,
b$ satisfying $(a+1)(b+1) = {\Omega} + 1$.  
\tdred{
For any rectangle $K = \{i_1, \ldots, i_a\} \times \{j_1, \ldots, j_b\}$, 
let $S_{K}$  denote a random state sequence with distribution $U^n({\bf s} | {\bf x}_{i_1}, \ldots, {\bf x}_{i_a}, {\bf y}_{j_1}, \ldots, 
{\bf y}_{j_b}) $.  For any rectangle $K'$ with $| I_{K'} | = a+1$ and 
$| J_{K'} | = b+1$ and any $(i, j) \in K'$, we let $K_{i,j}'$ denote the smaller rectangle $K_{i,j}' = I_{K'} \backslash \{ i \} \times 
J_{K'} \backslash \{ j \}$.  For any such $K'$ with one of its element being denoted by $(i^*, j^*)$, we have
}

\vspace{0.02in}
$\sum_{(i, j) \in K'}
    E[e_L(i, j, S_{K_{i,j}'})]$
\begin{eqnarray}
&=& \sum_{(i, j) \in K'} 
	\left ( 1 - 
	\hspace{-0.1in}
      \sum_{{\bf z}: (i, j) \in \phi_L({\bf z})}
      E[W^n({\bf z} | {\bf x}_{i}, {\bf y}_{j}, 
      S_{K_{i,j}'})] \right ) \nonumber
\end{eqnarray}
\begin{eqnarray}
&=& (a+1)(b+1) - \nonumber \\
& &   \sum_{{\bf z}}
      \mathop{\sum_{(i,j) \in K'}}_
      { (i, j) \in \phi_L({\bf z}) }
      E[W^n({\bf z} | {\bf x}_{i}, {\bf y}_{j}, 
      S_{K_{i,j}'})] \nonumber \\
&=& (a+1)(b+1) - \nonumber \\
& &   \sum_{{\bf z}}
\mathop{\sum_{(i,j) \in K'}}_
      { (i, j) \in \phi_L({\bf z}) }
      E[W^n({\bf z} | {\bf x}_{i^*}, {\bf y}_{j^*}, 
      S_{K_{i^*,j^*}'})] 
      \mbox{~~by~~} (\ref{sym-2})  \nonumber \\
&\geq& ({\Omega}+1) - L\ \ \ \ \mbox{by~}|\phi_L({\bf z})| \leq L.
\label{eqn:Proof-Thm2-2}
\end{eqnarray}
Next, let $\mathcal{P}_{(a,b)}$ be the set of all rectangles $K$ with 
$| I_K | = a$ and $| J_K | = b$.  Then, $|\mathcal{P}_{(a,b)}|
=\scriptsize{\left( \begin{array}{cc} M \\ a \end{array} \right)
\left( \begin{array}{cc} M \\ b \end{array} \right)}$ and\
\\

$\frac{1}{|\mathcal{P}_{(a,b)}|} \sum_{K \in \mathcal{P}_{(a,b)}}
    E[\bar{e}_L(S_{K})]$
\begin{eqnarray}
    &\geq&
    \frac{1}{|\mathcal{P}_{(a,b)}| M^2} \sum_{K \in \mathcal{P}_{(a,b)}}
        \sum_{(i, j) \in I_K^c \times J_K^c} E[e_L(i, j, S_K)] \nonumber \\
    &=&
    \frac{1}{|\mathcal{P}_{(a,b)}| M^2}
    \sum_{K' \in \mathcal{P}_{(a+1,b+1)}}
    \sum_{(i, j) \in K'} 
    E[e_L(i, j, S_{K_{i,j}'})] \nonumber \\
    &\geq&  \frac{|\mathcal{P}_{(a+1, b+1)}|({\Omega}+1)}
        {|\mathcal{P}_{(a,b)}| M^2}
    (1 - \frac{L}{{\Omega}+1}),\ \mbox{~by~}(\ref{eqn:Proof-Thm2-2}) \nonumber \\
    &=& \frac{(M-a)(M-b)}{M^2}(1-\frac{L}{{\Omega}+1}),\nonumber \\
    & & \mbox{by~} (a+1)(b+1)= {\Omega}+1.
        \nonumber
\end{eqnarray}
It now follows as in the previous case that $~int(C_L) =
\emptyset$. $\qed$

\vspace{0.05in} \textbf{Proof of Theorem 3:}  We start with some
standard notations.  For positive numbers $u, v$ and a collection
of sequences, each of length $n$, $({\bf x}_1, \ldots, {\bf x}_u,
{\bf y}_1, \ldots, {\bf y}_v, {\bf s}, {\bf z}) \in
{(\mathcal{X})}^{un}  \times
{(\mathcal{Y})}^{vn}  \times
\mathcal{S}^n \times \mathcal{Z}^n,\ P_{({\bf x}_1, \ldots, {\bf
x}_u, {\bf y}_1, \ldots, {\bf y}_v, {\bf s}, {\bf z})}$ denotes
the joint type of the sequences: the empirical distribution on
$\mathcal{X}^u \times \mathcal{Y}^v \times \mathcal{S} \times
\mathcal{Z}$ of the sequences which is given by the formula
\[
    \hspace{-0.1in} P_{({\bf x}_1, \ldots, {\bf x}_u, {\bf y}_1, \ldots, {\bf y}_v, {\bf s}, {\bf z})}(x_1, \ldots,         x_u, y_1, \ldots, y_v, s, z) \]
\[  \hspace{-0.3in} =  \frac{n_{(x_1, \ldots, x_u, y_1, \ldots, y_v, s, z)}}{n}, \]
where $n_{(x_1, \ldots, x_u, y_1, \ldots, y_v, s, z)}$ is the
number of $t \in \{1, \ldots, n\}$ such that $(x_{1t}, \ldots,
x_{ut}, y_{1t}, \ldots, y_{vt}, s_t, z_t) = (x_1, \ldots, x_u,
y_1, \ldots, y_v, s, z)$. Each of the marginal distributions of
$P_{({\bf x}_1, \ldots, {\bf x}_u, {\bf y}_1, \ldots, {\bf y}_v,
{\bf s}, {\bf z})}$ on $\mathcal{X}, \ldots, \mathcal{X},
\mathcal{Y}, \ldots, \mathcal{Y}, \mathcal{S}, \mathcal{Z}$ is
called the type of ${\bf x}_1, \ldots, {\bf x}_u, {\bf y}_1,
\ldots, {\bf y}_v, {\bf s}, {\bf z}$, respectively. Given the
joint type, it is often convenient to associate random variables
$X_1, \ldots, X_u, Y_1, \ldots, Y_v, S, Z$ with the joint
distribution $P_{({\bf x}_1, \ldots, {\bf x}_u, {\bf y}_1, \ldots,
{\bf y}_v, {\bf s}, {\bf z})}$.

For a finite set $\mathcal{A}$ and any two distributions $P_1(a),
P_2(a),\ a \in \mathcal{A}$, let $D(P_1 || P_2) \triangleq
\sum\limits_{a \in \mathcal{A}} P_1(a)
\log{\frac{P_1(a)}{P_2(a)}}$ and $d(P_1, P_2) \triangleq
\sum\limits_{a \in \mathcal{A}} |P_1(a) - P_2(a)|$ denote the
divergence and variational distance of $P_1, P_2$, respectively.

For a collection of sequences $({\bf x}_1, \ldots, {\bf x}_u, {\bf
y}_1, \ldots, {\bf y}_v, {\bf s}, {\bf z})$ with the joint type
$P_{({\bf x}_1, \ldots, {\bf x}_u, {\bf y}_1, \ldots, {\bf y}_v,
{\bf s}, {\bf z})} = P_{X^u Y^v S Z}$, we use the following
standard notations
\[\mathcal{T}_X = \{{\bf x} \in \mathcal{X}^n: P_{{\bf x}}
                        = P_X \} \]
\[\mathcal{T}_{Z|X_1 Y_1 S}({\bf x}_1, {\bf y}_1, {\bf s})
            = \{{\bf z} \in \mathcal{Z}^n:
                P_{({\bf x}_1, {\bf y}_1, {\bf s}, {\bf z})}
                = P_{X_1 Y_1 S Z} \} \]
\[\mathcal{T}_{Z|X^u Y^v S}({\bf x}^u, {\bf y}^v, {\bf s})
            = \{{\bf z} \in \mathcal{Z}^n:
                P_{({\bf x}^u, {\bf y}^v, {\bf s}, {\bf z})}
                = P_{X^u Y^v S Z} \} \]
\[\mathcal{T}_{X_1|X_2^u Y^v S}({\bf x}_2^u, {\bf y}^v, {\bf s})
            = \{{\bf x} \in \mathcal{X}^n:
                P_{({\bf x}, {\bf x}_2^u, {\bf y}^v, {\bf s})}
                = P_{X^u Y^v S} \}. \]
Then, the following relations are valid \cite{Csiszar_Korner_81}:

For any $W(z|x, y, s),$

\begin{eqnarray}
    W^n(\mathcal{T}_{Z|X_1 Y_1 S}({\bf x}_1, {\bf y}_1, {\bf s}) |{\bf x}_1, {\bf y}_1, {\bf
    s}) \nonumber \\
    \tdred{\leq}\ \ \ 2^{-n D(P_{X_1 Y_1 S Z} || W \times P_{X_1 Y_1 S})}; \label{eqn:types-1}
\end{eqnarray}
For any $W(z|x^u, y^v, s),$
\begin{eqnarray}
    W^n(\mathcal{T}_{Z|X^u Y^v S}({\bf x}^u, {\bf y}^v, {\bf s}) |{\bf x}_1, {\bf y}_1, {\bf s})
    \hspace{0.6in}
    \nonumber \\
    \tdred{\leq}\ \ \ 2^{-n D(P_{X^u Y^v S Z} || W(z| x_1, y_1, s) \times P_{X^u Y^v S})}; \label{eqn:types-2}
\end{eqnarray}
For any $Q(x_1|x_2^u, y^v, s),$
\begin{eqnarray}
    Q^n(\mathcal{T}_{X_1|X_2^u Y^v S}({\bf x}_2^u, {\bf y}^v, {\bf s}) |{\bf x}_2^u, {\bf y}^v, {\bf s})
    \nonumber \\
    \tdred{\leq}\ \ \ 2^{-n D(P_{X^u Y^v S} || Q \times P_{X_2^u Y^v S})}. \label{eqn:types-3}
\end{eqnarray}

Recall from Theorem 1 that it suffices to show that $int(C_L) \neq
\emptyset$.  To this end, we consider a ``constant composition''
code $\mathcal{U} = \mathcal{U}^{(n)}_{{\bf x}} = \{{\bf x}_1,
\ldots, {\bf x}_M\}$ and $\mathcal{V} = \mathcal{V}^{(n)}_{{\bf
y}} = \{{\bf y}_1, \ldots, {\bf y}_M\}$ with each ${\bf x}_i$ and
each ${\bf y}_j$ having the same types $P_X(x) = P_{({\bf x})}$ and
$P_Y(y) = P_{({\bf y})}$ coinciding with the types of fixed
sequences ${\bf x} \in \mathcal{X}^n$ and ${\bf y} \in
\mathcal{Y}^n$, respectively.  We first describe a list decoding
algorithm for such code and show in Lemma 1 that it is a
list-of-$L$ size decoder.  Then, a ``good'' code is specified in
Lemma 2 and is used, together with the decoder, to show that
$int(C_L) \neq \emptyset$.

The list decoding algorithm consists of two steps and
is parameterized by a (small) parameter $\eta > 0$ to be
chosen shortly.  This
algorithm follows the ideas of \cite{Csiszar_Narayan_88,
Blinovsky-Narayan-Pinsker_95, Hughes_97, Ahlswede-Cai_99}.

Given the received sequence ${\bf z}$, a successive execution of
the following two steps will produce the output list
$\phi_L^{\mathcal{U}, \mathcal{V}}({\bf z})$:

1.  Collect a list of message pairs $\Gamma \subseteq [M] \times
[M]$ comprising every $(i,j) \in \Gamma$ for which there exists a
state sequence ${\bf s} \in \mathcal{S}^n$ such that
\begin{equation}
    D(P_{X Y S Z} || P_{X} \times P_{Y} \times P_S \times W)
    \leq \eta,      \label{eqn:ProofThm3-1}
\end{equation}
where $P_{X Y S Z} = P_{({\bf x}_i, {\bf y}_j, {\bf s}, {\bf
z})}.$ If ${\bf z} \in \mathcal{Z}^n$ is such that $\vert \Gamma
\vert \leq L,$ then assign $\phi_L^{\mathcal{U}, \mathcal{V}}({\bf
z}) = \Gamma$ and stop.  For such ${\bf z},\ \Gamma$ is a
feasible decoded list of pairs of messages.  Otherwise, we proceed
to Step 2 to prune $\Gamma$ to get a feasible list as follows.

2.  Put a message pair $(i, j)$ in
$\phi_L^{\mathcal{U}, \mathcal{V}}({\bf z})$ if $(i,j) \in \Gamma$
and if for some ${\bf s} \in \mathcal{S}^n$ satisfying
(\ref{eqn:ProofThm3-1}), it holds that for every subset
$K \subseteq \Gamma$ such that $(i,j) \in K$ and $|K| = L+1$,
\begin{equation}
    I(X Y Z \wedge X^{a-1}, Y^{b-1} | S)
    \leq \eta,          \label{eqn:ProofThm3-2}
\end{equation}
where $a = |I_K|$ and $b = |J_K|$ and $P_{X Y X^{a-1} Y^{b-1} S Z}
= P_{({\bf x}_i, {\bf y}_j, {\bf x}_{I_{K} \backslash \{i\}},
{\bf y}_{J_{K} \backslash \{j\}}, {\bf s}, {\bf z})}$.

\vspace{0.05in}
Let $\phi_L^{\mathcal{U}, \mathcal{V}}({\bf z}) =
\{(1,1)\}$ if no $(i,j)$ satisfies (\ref{eqn:ProofThm3-1}) and
(\ref{eqn:ProofThm3-2}).

\vspace{0.05in}
\begin{lemma}
Let ${\Omega}$ be the
symmetrizability of an AVMAC.
Then, there exist 
\tdred{
functions
$f:\{0, 1, \ldots \} \rightarrow \{0, 1, \ldots \}$ and
$\eta(\alpha):\ \mathbb{R}^+ \rightarrow \mathbb{R}^+$
}
such that for any $\alpha > 0$, any ${\bf
x} \in \mathcal{X}^n$ and ${\bf y} \in \mathcal{Y}^n$ satisfying
$\min_{x \in \mathcal{X}} P_{({\bf x})}(x) \geq \alpha$ and
$\min_{y \in \mathcal{Y}} P_{({\bf y})}(y) \geq \alpha$,
every $L \geq f({\Omega})$, and any constant composition code
$(\mathcal{U}_{{\bf x}}, \mathcal{V}_{{\bf y}})$, the decoding
algorithm $\phi_L^{\mathcal{U}, \mathcal{V}} ({\bf z})$ as above and
with $\eta \left( \alpha \right)$ in (\ref{eqn:ProofThm3-1}) and
(\ref{eqn:ProofThm3-2})
satisfies $|\phi_L^{\mathcal{U}, \mathcal{V}} ({\bf z})| \leq L$
for every ${\bf z} \in \mathcal{Z}^n$. Furthermore, $f(u) \leq
(u+1)^2$ for every $u \geq 0$.
\end{lemma}

\textbf{Proof of Lemma 1:}  
For positive integers $A, M$, a set $K \subseteq [M] \times [M]$ is said to contain an $A$-diagonal or an $A$-rectangle if there exists $K' \subseteq K$
such that $K'$ is an $A$-diagonal or an $A$-rectangle,
respectively. 
For any positive integers $A, R, M$ such that $A
\leq R$, let $B(A, R, M) \triangleq |\{K \subseteq [M] \times [M]:
|K|=R,\ 
\tdred{
K \mbox{~contains~neither~}A
\mbox{-diagonal~nor~}A\mbox{-rectangle}
}
\}|$.

\vspace{0.05in}
{\em Claim 1: For any $A \geq 1,$ any $R \geq (A-1)^2 + 1$ and any $M
> 0,\ B(A, R, M) = 0$.}

To see this, consider an arbitrary $M > 0$
and a set $K \subseteq [M] \times [M]$ with $|K| = R \geq (A-1)^2
+1$. If $|I_K| \geq A$ and $|J_K| \geq A$, then clearly $K$
contains an $A$-diagonal.  Consider the rest of the $K'$s with,
say, $|I_K| \leq A - 1$.  As $(A-1)^2+1 \leq |K| = \sum_{i \in
I_K} |K \cap \{i\} \times J_K|$ and $|I_K| \leq A-1$, there exists
an $i^* \in I_K$ for which $|K \cap \{i^*\} \times J_K| \geq A$,
i.e., $K$ contains an $A$-rectangle, namely, $K \cap \{i^*\}
\times J_K$. This proves the claim.

Next, let
\begin{equation}
    g(A) \triangleq \mathop{\min_{R \geq A}}_{\sup_{M > 0} B(A, R, M) = 0} R \,\leq\, (A-1)^2 + 1,\ \mbox{by~Claim~1}.  \label{eqn:ProofThm3-3}
\end{equation}
The significance of $g$ in (\ref{eqn:ProofThm3-3}) can be understood as follows.
For $A \geq 1$ and {\em any} $M > 0$, it holds that
{\em any} set $K \subseteq [M] \times [M]$ with $|K| \geq g(A)$ must
contain either an $A$-diagonal or an $A$-rectangle.  We now let,
for every $u \geq 0$,
\begin{equation}
    f(u) \triangleq g(u+2) - 1 \leq (u+1)^2,\ \mbox{by}~
    (\ref{eqn:ProofThm3-3}). \label{eqn:ProofThm3-4}
\end{equation}
We shall prove Lemma 1 with this $f$ by
contradiction.  Suppose that there exists an output sequence
${\bf z} \in \mathcal{Z}^n$ such that $|\phi_L^{\mathcal{U}, \mathcal{V}}({\bf z})|
\geq L+1$.  Pick some $K \subseteq \phi_L^{\mathcal{U}, \mathcal{V}}({\bf z})$
with
\begin{equation}
    |K| = L + 1 \geq f({\Omega}) + 1 = g({\Omega}+2),\ \mbox{by~}(\ref{eqn:ProofThm3-4}).    \label{eqn:ProofThm3-5}
\end{equation}
Then, for any $(i, j) \in K$,
by (\ref{eqn:ProofThm3-1}) and (\ref{eqn:ProofThm3-2}),
we have that for some ${\bf s}_{ij} \in
\mathcal{S}^n$ with $P_{X_{I_K} Y_{J_K} S_{ij} Z} = P_{
({\bf x}_{I_K}, {\bf y}_{J_K}, {\bf s}_{ij}, {\bf z})}$, it holds that
\begin{eqnarray}
    2 \eta &\geq& D(P_{X_{i} Y_{j} S_{ij} Z} ||
                    P_{X_i} \times P_{Y_j} \times P_{S_{ij}} \times
                    W)\ \ \ \ \ \ \ \ \ \ \ \ \ \ \ \nonumber \\
            & &   +\ I(X_i Y_j Z \wedge X_{I_K \backslash \{i\}} Y_{J_K \backslash \{j\}} | S_{ij})
                  \nonumber \\
        &=& D \left(\begin{array}{ll}
                   P_{X_{I_K} Y_{J_K} S_{ij} Z} || \\
                   P_{X_i} \times P_{Y_j} \times P_{S_{ij} X_{I_K \backslash \{i\}} Y_{J_K \backslash \{j\}}} \times
                   W
              \end{array}\right ).
              \label{eqn:ProofThm3-5-2}
\end{eqnarray}
Next, from (\ref{eqn:ProofThm3-5}) and (\ref{eqn:ProofThm3-3}),
$K$ contains either a $({\Omega}+2)$-diagonal or a $({\Omega}+2)$-rectangle.

First, consider the case in which $K$ contains a $({\Omega}+2)$-diagonal.  Specifically, there exists a subset $K' \subseteq K$ such that $|K'|=|I_{K'}|=|J_{K'}|={\Omega}+2$.  By separately permuting the pair of
indices of $[M] \times [M]$, we can assume without any loss of
generality that $K' = \{(1,1), \ldots, ({\Omega}+2, {\Omega}+2)\}$.  Applying the
logsum inequality to
(\ref{eqn:ProofThm3-5-2}) to every $(i,i) \in K',\ i \in [{\Omega}+2]$,
we get \\
\begin{equation}
    2 \eta \geq D \left(\begin{array}{ll}
                   P_{X^{{\Omega}+2} Y^{{\Omega}+2} Z} || \\
                   P_{X_i} \times P_{Y_i} \times
                   (\sum_{s \in \mathcal{S}} 
                   P_{S_{ii} X^{{\Omega}+2}_{i} Y^{{\Omega}+2}_{i}} \times
                   W)
              \end{array} \right),
              \label{eqn:ProofThm3-6-1}
\end{equation}
where $X^{{\Omega}+2}_{i} \triangleq X_{[{\Omega}+2] \backslash \{i\}}$ and $Y^{{\Omega}+2}_{i} \triangleq Y_{[{\Omega}+2] \backslash \{i\}}$.  Applying Pinsker's inequality \cite[p. 58]{Csiszar_Korner_81} to
(\ref{eqn:ProofThm3-6-1}), we get that,
for each $i \in [{\Omega}+2]$,
\begin{equation}
             c \sqrt{2 \eta} \geq d \left(
                \begin{array}{ll}
                   P_{X^{{\Omega}+2} Y^{{\Omega}+2} Z}, \\
                   P_{X_i} \times P_{Y_i} \times
                   (\sum_{s \in \mathcal{S}}
                   P_{S_{ii} X^{{\Omega}+2}_i Y^{{\Omega}+2}_i} \times
                   W)
              \end{array} \right),
              \label{eqn:ProofThm3-6}
\end{equation}
where $c$ is an absolute constant.
With triangle inequality, we obtain that

\vspace{0.05in}
$\hspace{-0.12in} 2c \sqrt{2 \eta} \ \ \geq$
\begin{align}
			\max_{1 \leq i < j \leq {\Omega}+2}
                  \hspace{-0.05in} 
              	 d \left( \hspace{-0.09in}
                   \begin{array}{ll}
                   P_{X_i} \times P_{Y_i} \times
                   (\sum_{s \in \mathcal{S}} 
                   P_{S_{ii} X^{{\Omega}+2}_i Y^{{\Omega}+2}_{i}} \times
                   W), \\
                   P_{X_j} \times P_{Y_j} \times
                   (\sum_{s \in \mathcal{S}} 
                   P_{S_{jj} X^{{\Omega}+2}_j Y^{{\Omega}+2}_{j}} \times
                   W)
              	\end{array} 
              	\hspace{-0.09in}
              	\right). \nonumber \\
              \label{eqn:ProofThm3-7}
\end{align}
Note that $P_{X_i}=P_{({\bf x})}$ and $P_{Y_i}=P_{({\bf y})}$, $i
= 1, \ldots, {\Omega}+2$, with $\min_{x \in \mathcal{X}} P_{({\bf x})}(x)
\geq \alpha$ and $\min_{y \in \mathcal{Y}} P_{({\bf y})}(y) \geq
\alpha$, respectively.  The sought contradiction is obtained by
invoking the following Claim 2 upon setting $\eta$ sufficiently
small.  The proof of Claim 2 is similar to that of Lemma A4 of
\cite{Hughes_97} and is relegated to the Appendix A.

\vspace{0.05in} {\em Claim 2: For an AVMAC $T$ with
symmetrizability ${\Omega}$ and any $\alpha > 0$, there exists
$\nu(\alpha) > 0$ such that for any pair of distributions $P(x),\
x \in \mathcal{X}$, and $Q(y),\ y \in \mathcal{Y}$, satisfying
$\min_{x \in \mathcal{X}} P(x) \geq \alpha$, $\min_{y \in
\mathcal{Y}} Q(y) \geq \alpha$ and any collection of ${\Omega}+2$
\tdred{joint distributions $U_i$ on 
$\mathcal{X}^{{\Omega}+1} \times \mathcal{Y}^{{\Omega}+1} \times
\mathcal{S},\ i = 1, \ldots, {\Omega}+2$, it holds that}
\tdred{
\begin{eqnarray}
\max_{1 \leq i < j \leq {\Omega}+2}\ \  d \left (
    P_{X^{{\Omega}+2} Y^{{\Omega}+2} S}^{(i)} \,,\  
    P_{X^{{\Omega}+2} Y^{{\Omega}+2} S}^{(j)}
\right ) \ \ \geq\ \  \nu, \label{eqn:Lemma_2}
\end{eqnarray}
where for $i = 1, \ldots, {\Omega}+2,$ the joint distribution
$P_{X^{{\Omega}+2} Y^{{\Omega}+2} S}^{(i)}$ on 
$\mathcal{X}^{{\Omega}+2} \times \mathcal{Y}^{{\Omega}+2} \times \mathcal{S}$ is
\begin{align}
P_{X^{{\Omega}+2} Y^{{\Omega}+2} S}^{(i)}(x^{{\Omega}+2}, y^{{\Omega}+2}, s) 
\hspace{1.7in}
\nonumber \\ 
=\ \ 
P(x_i)Q(y_i) 
\left( \sum_{s \in \mathcal{S}} W(z|x_i, y_i, s)
U_i(x^{{\Omega}+2}_{i}, y^{{\Omega}+2}_{i}, s) \right). \nonumber
\end{align}
}
}

Lastly, we consider the case in which $K$ contains a
$({\Omega}+2)$-rectangle. Precisely, there exists $I \times J \subseteq
K$ with $|I| = a+1$, $|J| = b+1$ and $(a+1)(b+1) \geq {\Omega}+2$.  By
separately permuting the pair of indices of $[M] \times [M]$, we
can assume without any loss of generality that $I=[a+1],\
J=[b+1]$. Similar to the argument leading to
(\ref{eqn:ProofThm3-7}), we get

\vspace{0.05in}
$\hspace{-0.17in} 2c \sqrt{2 \eta} \ \ \geq$
\begin{align}
\max_{(i, j), (i', j') \in [a+1] \times [b+1],\ (i, j) \neq (i', j')} \hspace{1.2in} \nonumber \\
              \hspace{0.03in} 
              d \left(
              \hspace{-0.07in} 
              \begin{array}{ll}
                   P_{X_i} \times P_{Y_j} \times
                   (\sum_{s \in \mathcal{S}} P_{S_{ij} X^{a+1}_i Y^{b+1}_j} \times
                   W), \\
                   P_{X_{i'}} \times P_{Y_{j'}} \times
                   (\sum_{s \in \mathcal{S}} P_{S_{i'j'} X^{a+1}_{i'} Y^{b+1}_{j'}} \times
                   W)
              \end{array} 
              \hspace{-0.09in} 
              \right).
              \label{eqn:ProofThm3-8}
\end{align}
The sought contradiction is obtained by invoking the following
Claim 3, whose proof is also given in Appendix A, upon setting
$\eta$ sufficiently small.

\vspace{0.05in} {\em Claim 3: For an AVMAC $T$ with
symmetrizability ${\Omega}$ and any $\alpha > 0$, there exists
$\nu(\alpha) > 0$ such that for any pair of distributions $P(x),\
x \in \mathcal{X}$, and $Q(y),\ y \in \mathcal{Y}$, satisfying
$\min_{x \in \mathcal{X}} P(x) \geq \alpha$, $\min_{y \in
\mathcal{Y}} Q(y) \geq \alpha$ and any collection of $(a+1)(b+1)
\geq {\Omega}+2$ joint distributions 
\tdred{
$U_{ij}$ on $\mathcal{X}^{a} \times \mathcal{Y}^{b} \times \mathcal{S},$\\
$i = 1, \ldots, a+1,\  j = 1 \ldots, b+1$, it holds that}
\tdred{
\begin{align}
\mathop{\max_{(i,j), (i', j') \in [a+1] \times [b+1] }}_{(i,j) \neq (i', j')}
d \left (
\hspace{-0.02in}
    P_{X^{a+1} Y^{b+1} S}^{(i j)} \,, 
    P_{X^{a+1} Y^{b+1} S}^{(i' j')}
\hspace{-0.02in}
\right ) \,\geq\,   \nu,
\label{eqn:Lemma_3}
\end{align}
where for $i = 1, \ldots, a+1,\ j = 1 \ldots, b+1,$ the joint distribution
$P_{X^{a+1} Y^{b+1} S}^{(i j)}$ on $\mathcal{X}^{a+1} \times \mathcal{Y}^{b+1} \times \mathcal{S}$ is
\begin{align}
P_{X^{a+1} Y^{b+1} S}^{(i j)} (x^{a+1}, y^{b+1}, s) 
\hspace{1.7in}
\nonumber \\ 
=\ 
P(x_i)Q(y_j)
    \left( \sum_{s \in \mathcal{S}}
    W(z|x_i, y_j, s) U_{ij}(x^{a+1}_i, y^{b+1}_j, s) \right). \nonumber
\end{align}
}
}
This completes the proof of Lemma 1. $\qed$

We now specify in the following Lemma 2 a ``good'' deterministic
code, with nonzero rates; the proof of the lemma is similar to
that of Lemma 2 in \cite{Ahlswede-Cai_99} and is relegated to
Appendix B.

For a deterministic code $(\mathcal{U}^{(n)}_{{\bf x}},
\mathcal{V}^{(n)}_{{\bf y}})$ with $|\mathcal{U}| = |\mathcal{V}|
= M$ and $R = \frac{1}{n}\log_2{M}$, any $\epsilon > 0$ and any
${\bf s} \in \mathcal{S}^n$, we define
\begin{eqnarray}
    \mathcal{A}_{\epsilon}({\bf s}) \triangleq
    \left \{
    \begin{array}{ll}
        (i,j) \in [M] \times [M]: \\
        D(P_{X Y S}||P_X \times P_Y \times P_S) < \epsilon\\
        \mbox{where~}P_{XYS} = P_{({\bf x}_i, {\bf y}_j, {\bf s})}
    \end{array}
    \right \}, \label{eqn:ProofThm3-9}
\end{eqnarray}
\begin{eqnarray}
    \mathcal{B}_{\epsilon}({\bf s}) \triangleq
    \left \{
    \begin{array}{ll}
        i \in [M]: \mbox{for~any~}I \subseteq [M] \backslash \{i\}, \\
        |I| = L, \mbox{~any~} J \subseteq [M],\ |J| = L + 1,\ \\
        I(X \wedge X_I, Y_J, S) < (2L + 1)R + \epsilon,\ \\
        \mbox{where}\ P_{X X_I Y_J S} =
        P_{({\bf x}_i, {\bf x}_I, {\bf y}_J, {\bf s})}
    \end{array}
    \right \},     \label{eqn:ProofThm3-10}
\end{eqnarray}
\begin{eqnarray}
   \mathcal{C}_{\epsilon}({\bf s}) \triangleq
    \left \{
    \begin{array}{ll}
        j \in [M]: \mbox{for~any~}J \subseteq [M] \backslash \{j\}, \\
        |J| = L, \mbox{~any~} I \subseteq [M],\ |I| = L + 1,\ \\
        I(Y \wedge X_I, Y_J, S) < (2L + 1)R + \epsilon,\ \\
        \mbox{where}\ P_{X_I Y Y_J S} =
        P_{({\bf x}_I, {\bf y}_j, {\bf y}_J, {\bf s})}
    \end{array}
    \right \},     \label{eqn:ProofThm3-11}
\end{eqnarray}


\begin{lemma}
For any $0 < \epsilon < \delta$, \tdred{all sufficiently large $n$, and} 
any sequences ${\bf x} \in \mathcal{X}^n$ and ${\bf y}
\in \mathcal{Y}^n$ with $H(P_{({\bf x})}) > \delta$ and
$H(P_{({\bf y})}) > \delta$, there
exists a deterministic code $(\mathcal{U}_{{\bf x}},
\mathcal{V}_{{\bf y}})$ as above with $R \geq \delta$ such that
for every ${\bf s} \in \mathcal{S}^n$,
\begin{equation}
    |\mathcal{A}_{\epsilon}({\bf s})^c| \leq 2^{-\frac{\epsilon}{4}n}M^2\ \mbox{and}
    \label{eqn:ProofThm3-12}
\end{equation}
\begin{equation}
    |\mathcal{B}_{\epsilon}({\bf s})^c|, |\mathcal{C}_{\epsilon}({\bf s})^c| \leq 2^{-\frac{\epsilon}{4}n}M.
    \label{eqn:ProofThm3-13}
\end{equation}
\end{lemma}
%
For a fixed $\alpha > 0$ and all $n$ sufficiently large, choose
${\bf x} \in \mathcal{X}^n$ and ${\bf y} \in \mathcal{Y}^n$ so
that $\min_{x \in \mathcal{X}} P_{({\bf x})}(x) \geq \alpha$ and
$\min_{y \in \mathcal{Y}} P_{({\bf y})}(y) \geq \alpha$.  We then
choose $\eta$ sufficiently small according to Lemma 1.  Next, for
the ${\bf x}$ and ${\bf y}$, and for some $\epsilon$ and $\delta$
sufficiently small so that $H(P_{({\bf x})}) > \delta$ and
$H(P_{({\bf y})}) > \delta$ and
\begin{equation}
    0 < \epsilon < \delta \leq R < \frac{\eta}{2(6L+4)},     \label{eqn:ProofThm3-14}
\end{equation}
we get from Lemma 2 a deterministic code $(\mathcal{U}_{{\bf x}},
\mathcal{V}_{{\bf y}})$ satisfying (\ref{eqn:ProofThm3-12}),
(\ref{eqn:ProofThm3-13}) and (\ref{eqn:ProofThm3-14}) with rate
$R$. Combining this code with the decoding algorithm from Lemma 1,
we obtain a deterministic code decoded into a list of size $L$.
Lastly, we show that for every ${\bf s} \in \mathcal{S}^n$,
$\bar{e}_L({\bf s})$ approaches zero exponentially fast.

First, we note that it suffices to prove that for all $(i, j) \in
\mathcal{A}_{\epsilon}({\bf s}) \cap [\mathcal{B}_{\epsilon}({\bf s})
\times \mathcal{C}_{\epsilon}({\bf s})],\ e_{L}(i,j, {\bf s})$
approaches zero exponentially fast, because by (\ref{eqn:ProofThm3-12})
and (\ref{eqn:ProofThm3-13}),
\begin{equation}
    \bar{e}_L({\bf s}) \leq
    \frac{1}{M^2} \sum_{(i,j) \in \mathcal{A}_{\epsilon}({\bf s}) \cap [\mathcal{B}_{\epsilon}({\bf s})
    \times \mathcal{C}_{\epsilon}({\bf s})]} e_L(i,j, {\bf s}) +
    3 \times 2^{-\frac{\epsilon}{4}n}.
\nonumber
\end{equation}
For a fixed ${\bf s}$ and $(i, j) \in \mathcal{A}_{\epsilon}({\bf
s}) \cap [\mathcal{B}_{\epsilon}({\bf s}) \times
\mathcal{C}_{\epsilon}({\bf s})],\ e_{L}(i,j, {\bf s})$ is upper
bounded by the probability of the event
\begin{equation}
    \bigcap_{{\bf s}' \in \mathcal{S}^n} \left \{ E_0({\bf s}') \bigcup
    \left (\bigcup_{(a,b): \mathcal{K}_{a,b}^{L, (i,j)} \neq \emptyset } E_{a,b}({\bf s}') \right ) \right \},
    \label{eqn:ProofThm3-15}
\end{equation}
with respect to the conditional probability distribution $W^n({\bf
z} | {\bf x}_i, {\bf y}_j, {\bf s})$.  In
(\ref{eqn:ProofThm3-15}), $E_0({\bf s}')$ is the set of all ${\bf
z} \in \mathcal{Z}^n$ for which (\ref{eqn:ProofThm3-1}) is
violated with ${\bf x}_i, {\bf y}_j, {\bf s}', {\bf z}$ and each
of the $E_{a,b}({\bf s}')$ is the set of all ${\bf z} \in
\mathcal{Z}^n$ for which (\ref{eqn:ProofThm3-2}) is violated with
${\bf x}_i, {\bf y}_j, {\bf x}_{I_K \backslash \{i\}}, {\bf
y}_{J_K \backslash \{j\}}, {\bf s}', {\bf z}$ for some $K \in
\mathcal{K}_{a,b}^{L, (i,j)}$, where $\mathcal{K}_{a,b}^{L, (i,j)}
\triangleq \{ K \subseteq [M] \times [M],\ (i, j) \in K,\ |K| =
L+1,\ |I_{K}| = a,\ |J_{K}| = b\}$.  As
\begin{equation}
    E_0({\bf s}) \bigcup \left ( \bigcup_{(a,b): \mathcal{K}_{a,b}^{L, (i,j)} \neq \emptyset } E_{a,b}({\bf s}) \right)
    \label{eqn:ProofThm3-15-b}
\end{equation}
subsumes (\ref{eqn:ProofThm3-15}), it suffices to prove the
exponential decays of $W^n(E_0({\bf s}))|{\bf x}_i, {\bf y}_j,
{\bf s})$ and $W^n(E_{a, b}({\bf s}))|{\bf x}_i, {\bf y}_j, {\bf
s})$, for every ${\bf s}$, $(i, j) \in \mathcal{A}_{\epsilon}({\bf
s}) \cap [\mathcal{B}_{\epsilon}({\bf s}) \times
\mathcal{C}_{\epsilon}({\bf s})]$ and $(a, b)$ such that
$\mathcal{K}_{a,b}^{L, (i,j)} \neq \emptyset,$ as the number of
all possible such pairs $(a , b)$ is upper bounded by $(L+1)^2$.
To this end, it is convenient to  let
\begin{equation}
    p(n) \triangleq (n+1)^{|\mathcal{X}|^{L+1}|\mathcal{Y}|^{L+1}|\mathcal{S}||\mathcal{Z}|}.
    \label{eqn:ProofThm3-16}
\end{equation}
We start with $E_0({\bf s})$.  First, because $(i, j) \in
\mathcal{A}_{\epsilon}({\bf s})$, we get from
(\ref{eqn:ProofThm3-9}) that
\begin{equation}
    D(P_{XYS} || P_X \times P_Y \times P_S) \leq \epsilon,
    \label{eqn:ProofThm3-17}
\end{equation}
where $P_{XYS} = P_{({\bf x}_i, {\bf y}_j, {\bf s})}$. Next, we
let,
\begin{equation}
    \mathcal{Q}_0 \triangleq
    \left \{
    \begin{array}{ll}
        P_{X Y S Z}: P_{X Y S Z} = P_{({\bf x}_i, {\bf y}_j, {\bf s}, {\bf z})} \\
        \mbox{~for~some~} {\bf z} \in \mathcal{Z}^n \mbox{~such~that} \\
        D(P_{XYSZ} || P_X \times P_Y \times P_S \times W) \geq \eta
    \end{array}
    \right \}.
    \label{eqn:ProofThm3-18}
\end{equation}
Then,\\

$\hspace{-0.1in} W^n(E_0({\bf s})| {\bf x}_i, {\bf y}_j , {\bf
s})$
\begin{align}
    &=\ \sum_{(X, Y, S, Z) \in \mathcal{Q}_0}
        W^n(\mathcal{T}_{Z|XYS}({\bf x}_i, {\bf y}_j , {\bf s})| {\bf x}_i, {\bf y}_j , {\bf s}) \nonumber \\
    &\leq\ p(n) \max_{(X, Y, S, Z) \in \mathcal{Q}_0}
           2^{-nD(P_{ZXYS}||W \times P_{XYS})}\ \mbox{by~}(\ref{eqn:types-1}) \nonumber \\
    &\leq\ p(n) \times \nonumber \\
    &  \  \max_{(X, Y, S, Z) \in \mathcal{Q}_0}  2^{ - n \left ( 
    	      \hspace{-0.08in}
            \begin{array}{ll}
                D(P_{ZXYS}||W \times P_{XYS}) - \epsilon \\
                + D(P_{XYS}|| P_{X} \times P_Y \times P_S)
            \end{array}
            \hspace{-0.08in}
            \right )
           }, \mbox{by~(\ref{eqn:ProofThm3-17})} \nonumber
\end{align}
\begin{align}
    &=\ p(n) \max_{(X, Y, S, Z) \in \mathcal{Q}_0}
          2^{ - n(D(P_{ZXYS}||P_{X} \times P_Y \times P_S \times W) - \epsilon)}
        \nonumber \\
    &\leq\ p(n) 2^{-n(\eta - \epsilon)}
    \ \leq\   p(n) 2^{-n(\frac{\eta}{2})},
    \ \mbox{~by~}(\ref{eqn:ProofThm3-18})~\mbox{and}~(\ref{eqn:ProofThm3-14}). \nonumber
\end{align}
Lastly, we tackle $E_{a, b}({\bf s})$ for each fixed $(a, b)$ for
which $\mathcal{K}_{a,b}^{L, (i,j)} \neq \emptyset$. Let
\begin{equation}
    \mathcal{Q}_{a,b} \triangleq
    \left \{
    \begin{array}{ll}
        P_{X Y X^{a-1} Y^{b-1} Z S}: P_{X Y X^{a-1} Y^{b-1} Z S} = \\
        P_{({\bf x}_i, {\bf y}_j, {\bf x}_{I_K \backslash \{i\}}, {\bf y}_{J_K \backslash \{j\}}, {\bf s}, {\bf z})}, \\
        \mbox{~for~some~}{\bf z}\mbox{~and~}K \subseteq [M] \times [M] \mbox{~such~that} \\
        (i,j) \in K,\ |K| = L+1,\ |I_K| = a,\ |J_K| = b, \\
        I(X Y Z \wedge X^{a-1} Y^{b-1}|S) \geq \eta
    \end{array}
    \right \}
    \label{eqn:ProofThm3-19}
\end{equation}
and\\

$\mathcal{R}_{a, b}(X Y X^{a-1} Y^{b-1} S) \triangleq $
\begin{equation}
    \left \{
    \begin{array}{ll}
    K \subseteq [M] \times [M]:\ (i,j) \in K,\\
    |K| = L+1,\ |I_K| = a, |J_K| = b, \\
    P_{({\bf x}_i, {\bf y}_j, {\bf x}_{I_K \backslash \{i\}}, {\bf y}_{J_K \backslash \{j\}}, {\bf s}, {\bf z})}
    = P_{X Y X^{a-1} Y^{b-1} S}
    \end{array}
    \right \}.
\end{equation}
Note that $|\mathcal{R}_{a, b}(X Y X^{a-1} Y^{b-1} S)| \leq
(M^2)^L$.

By the definition of $\mathcal{B}_{\epsilon}({\bf s})$ and
$\mathcal{C}_{\epsilon}({\bf s})$, for any joint type $X Y X^{a-1}
Y^{b-1} Z S$ in $\mathcal{Q}_{a, b}$, we get from
(\ref{eqn:ProofThm3-10}) and (\ref{eqn:ProofThm3-11}) that $I(X
\wedge X^{a-1} Y Y^{b-1} S) \leq (2L + 1)R + \epsilon$ and \\ $I(Y
\wedge X X^{a-1} Y^{b-1} S) \leq (2L + 1)R + \epsilon$, which
gives
\begin{eqnarray}
I(X Y \wedge X^{a-1} Y^{b-1}| S) &=&  I(X \wedge X^{a-1} Y^{b-1}| S) + \nonumber \\
& & I(Y \wedge X^{a-1} Y^{b-1}| X, S) \nonumber \\
    &\leq& 2(2L+1)R + 2\epsilon.
\label{eqn:ProofThm3-20}
\end{eqnarray}

Then,

$\hspace{-0.02in} W^n(E_{a, b}({\bf s}) | {\bf x}_i, {\bf y}_j, {\bf s})$
\begin{eqnarray}
    &\leq& \cup_{X Y X^{a-1} Y^{b-1} Z S \in \mathcal{Q}_{a, b}} \nonumber \\
    & &    \cup_{K \in \mathcal{R}_{a, b}(X Y X^{a-1} Y^{b-1} S)} \nonumber \\
    & & \hspace{-0.2in} W^n
           \left (
           \begin{array}{ll}
           \mathcal{T}_{Z|X Y X^{a-1} Y^{b-1} Z S} ({\bf x}_i, {\bf y}_j,
           {\bf x}_{I_K \backslash \{i\}}, {\bf y}_{J_K \backslash \{j\}},
           {\bf s} ) \\
           ~\big \vert ~{\bf x}_i, {\bf y}_j, {\bf s}
           \end{array}
           \right ) \nonumber \\
    &\leq& p(n) 2^{2L R} \times \nonumber \\
    & & \hspace{-0.15in}
    \max_{X Y X^{a-1} Y^{b-1} Z S \in \mathcal{Q}_{a,b}}
    \hspace{-0.18in}
    2^{-n D(P_{X X^{a-1} Y Y^{b-1} S Z} || W \times P_{X X^{a-1} Y Y^{b-1} S}) }, \nonumber \\
    & &     \mbox{by~}(\ref{eqn:types-2}) \nonumber
\end{eqnarray}
\begin{eqnarray}
    &=& p(n) 2^{2L R} \times \nonumber \\
    & &  \hspace{-0.05in}
    \max_{X Y X^{a-1} Y^{b-1} Z S \in \mathcal{Q}_{a,b}}
    \hspace{-0.32in}
    2^{-n
    \left(
    \begin{array}{cc}
    I(Z \wedge X^{a-1} Y^{b-1} | X Y S) +\\
    D(P_{X Y S Z} || W \times P_{X Y S})
    \end{array}
    \right)
    }, \nonumber \\
    &\leq& p(n) 2^{2L R}
        \hspace{-0.05in}
        \max_{X Y X^{a-1} Y^{b-1} Z S \in \mathcal{Q}_{a, b}}
    \hspace{-0.05in}
            2^{-n I(Z \wedge X^{a-1} Y^{b-1} | X Y S)} \nonumber \\
    &\leq& p(n) 2^{2L R} \times \nonumber \\
    & & \hspace{-0.1in} \max_{X Y X^{a-1} Y^{b-1} Z S \in \mathcal{Q}_{a, b}}
            2^{-n \left (
                \begin{array}{ll}
                I(X Y Z \wedge X^{a-1} Y^{b-1} | S) \\
                - 2(2L+1)R - 2 \epsilon
                \end{array}
                \right )
                }, \nonumber \\
    & & \mbox{by~}(\ref{eqn:ProofThm3-20}) \nonumber \\
    &\leq& p(n)
            2^{-n(\eta - (6L+2)R - 2 \epsilon)},\
            \mbox{by~}(\ref{eqn:ProofThm3-19}) \nonumber \\
    &\leq& p(n) 2^{-n(\frac{\eta}{2})},\
            \mbox{by~}(\ref{eqn:ProofThm3-14}).
            \nonumber
\end{eqnarray}
This completes the proof of Theorem 3. $\qed$

\vspace{0.05in} \textbf{Proof of Theorem 4:}  For any $x, y \in
\left \{ 0, 1 \right \},$ we let $W_{xy} = \left [ {\bf w}_{xy}^0,
{\bf w}_{xy}^1 \right ]$ denote the line segment on the simplex in
$\mathbb{R}^2:\ \left \{ \left( x, y \right):\ x \geq 0,\ y \geq
0,\ x + y = 1 \right \}$ connecting the two points ${\bf w}_{xy}^0
= \left( W \left( 0 \vert x, y , 0 \right), W \left( 1 \vert x, y
, 0 \right) \right)$ and ${\bf w}_{xy}^1 = \left( W \left( 0 \vert
x, y , 1 \right), W \left( 1 \vert x, y , 1 \right) \right).$
Furthermore, for any pmf $q \left( \cdot \right)$ on $\left \{ 0,
1 \right \},$ we let ${\bf w}_{xy}^q$ denote the point ${\bf
w}_{xy}^q = q \left( 0 \right) {\bf w}_{xy}^0 + q \left( 1 \right)
{\bf w}_{xy}^1.$

First, it suffices to assume that
\begin{equation}
\nexists q^*:\ {\bf w}_{00}^{q^*} \ =\ {\bf w}_{10}^{q^*}\ \ \mbox{and}\ \
{\bf w}_{01}^{q^*} \ =\ {\bf w}_{11}^{q^*};
\label{eqn-Pf-Thm4-zero-rand-cap-1}
\end{equation}
and
\begin{equation}
\nexists q^*:\  {\bf w}_{00}^{q^*} \ =\ {\bf w}_{01}^{q^*}\ \
\mbox{and}\ \ {\bf w}_{10}^{q^*} \ =\ {\bf w}_{11}^{q^*},
\label{eqn-Pf-Thm4-zero-rand-cap-2}
\end{equation}
otherwise (\ref{eqn-Pf-Thm4-zero-rand-cap-1}) or
(\ref{eqn-Pf-Thm4-zero-rand-cap-2}) will imply (see (\ref{eqn-rand-cap})) that
$int \left( C^R \right) = \emptyset$.

\vspace{-0.3in}
\begin{figure}[h]
\begin{center}
\setlength{\unitlength}{1bp}%
\begin{picture}(100,120)(0,0)
\put(20,80){\circle*{3}} \put(19.8,79.8){\line(1,-1){40}}
\put(35,65){\circle*{3}} \put(60,40){\circle*{3}}
\put(10,10){\line(1,0){80}} \put(10,10){\line(0,1){80}}
\put(10,90){\line(1,-1){80}} \put(22, 82){${\bf w}_{xy}^0$}
\put(27, 47){$W_{xy}$} \put(62,42){${\bf w}_{xy}^1$}
\put(37,67){${\bf w}_{xy}^{q^*}$}
\end{picture}
\end{center}
\end{figure}

Next, we observe that if $W_{00} \cap W_{10} = \emptyset
~\mbox{or}~ W_{01} \cap W_{11}$\\ $= \emptyset ~\left( W_{00} \cap
W_{01} = \emptyset ~\mbox{or}~ W_{10} \cap W_{11} = \emptyset
\right),$ then there is neither $U \left( s \vert x_2, y_2,
\ldots, x_{u+1}, y_{u+1} \right),\ u \geq 1,$ that fulfills
(\ref{sym-1}) nor $U \left( s \vert x_2, \ldots, x_{a+1}, y_2,
\ldots, y_{b+1} \right),\ a \geq 1 ~\left( b \geq 1\right),$ that
fulfills (\ref{sym-2}), respectively.  In order to establish the
finiteness of the symmetrizability ${\Omega},$ we first show the
following claim:

{\em Claim 4:  If $W_{00} \cap W_{10} \neq \emptyset ~\mbox{and}~
W_{01} \cap W_{11} \neq \emptyset $\\ $\left( W_{00} \cap W_{01}
\neq \emptyset ~\mbox{and}~ W_{10} \cap W_{11} \neq \emptyset
\right),$ then under (\ref{eqn-Pf-Thm4-zero-rand-cap-1})
((\ref{eqn-Pf-Thm4-zero-rand-cap-2})), there are finitely
many $u \geq 1$ for which there exist $U \left( s \vert x_2, y_2,
\ldots, x_{u+1}, y_{u+1} \right)$ that fulfill (\ref{sym-1}), and
there are {\em finitely many} $a \geq 1$ ($b \geq 1$) for which
there exist $U \left(s \vert x_2, \ldots, x_{a+1}, y_2, \ldots,
y_{b+1} \right)$ that fulfill (\ref{sym-2}).}

Note that the finiteness of the symmetrizability ${\Omega}$ will follow
from the claim (along with its symmetric version with the
bracketed statements) along with the previously mentioned
nonexistence of the $U \left( s \vert x_2, y_2, \ldots, x_{u+1},
y_{u+1} \right),\ u \geq 1,$ or $U \left( s \vert x_2, \ldots,
x_{a+1}, y_2, \ldots, y_{b+1} \right),\ a \geq 1 \left( b \geq 1
\right)$ under the condition $W_{00} \cap W_{10} = \emptyset
~\mbox{or}~ W_{01} \cap W_{11} = \emptyset$\\ $\left( W_{00} \cap
W_{01} = \emptyset ~\mbox{or}~ W_{10} \cap W_{11} = \emptyset
\right),$ respectively.

It is clear that we only need to establish the non-bracketed
version of Claim 4, as the bracketed version will follow by
symmetry. To this end, we assume that $W_{00} \cap W_{10} \neq
\emptyset,\ W_{01} \cap W_{11} \neq \emptyset$~and
(\ref{eqn-Pf-Thm4-zero-rand-cap-1}), and first prove that any $U
\left( s \vert x_2, \ldots, x_{a+1}, y_2, \ldots, y_{b+1} \right)$
that fulfills (\ref{sym-2}) must satisfy $a \leq K,$ where $K$ is
a constant depending only on the AVMAC.

Fix the sequence $\left( y_2, \ldots, y_{b+1} \right),$ and let
$\alpha_k = U \left( 1 \vert 1^k, 0^{a-k}, y_2, \ldots, y_{b+1} \right),\
k = 0, 1, \ldots, a,$ where $\left(x_2, \ldots, x_{a+1} \right) = \left( 1^k, 0^{a-k} \right),$ and $1^k, 0^{a-k}$ denote a string of 1 of length $k$ and a string of 0 of length $a-k,$ respectively.  Then if we denote the pmf $W \left( \cdot \vert x, y, s \right)$ by a row vector $\bf{w}_{xy}^s,$
then we conclude from (\ref{sym-2}) that for any $k = 0, 1,
\ldots, a-1,$
\begin{eqnarray}
\left [
\begin{array}{cc}
\bf{w}_{10}^0, \bf{w}_{10}^1\\
\bf{w}_{11}^0, \bf{w}_{11}^1
\end{array}
\right ]
\left [
 \begin{array}{cc}
 1 - \alpha_k \\
 \alpha_k
 \end{array}
\right ] =
\left [
\begin{array}{cc}
\bf{w}_{00}^0, \bf{w}_{00}^1\\
\bf{w}_{01}^0, \bf{w}_{01}^1
\end{array}
\right ]
\left [
 \begin{array}{cc}
 1 - \alpha_{k+1} \\
 \alpha_{k+1}
 \end{array}
\right ].
\label{eqn-Pf-Thm4-region-R2-increment}
\end{eqnarray}
It is clear that the set of all pairs $\left( \alpha, \tilde{\alpha} \right) \in
\left [0, 1 \right]^2$ that satisfy
\begin{eqnarray}
\left [
\begin{array}{cc}
\bf{w}_{10}^0, \bf{w}_{10}^1\\
\bf{w}_{11}^0, \bf{w}_{11}^1
\end{array}
\right ]
\left [
 \begin{array}{cc}
 1 - \alpha \\
 \alpha
 \end{array}
\right ] =
\left [
\begin{array}{cc}
\bf{w}_{00}^0, \bf{w}_{00}^1\\
\bf{w}_{01}^0, \bf{w}_{01}^1
\end{array}
\right ]
\left [
 \begin{array}{cc}
 1 - \tilde{\alpha} \\
 \tilde{\alpha}
 \end{array}
\right ]
\label{eqn-Pf-Thm4-region-R2}
\end{eqnarray}
is closed and convex.  We denote this set by $S(W) \in \left [ 0, 1 \right ]^2.$
From (\ref{eqn-Pf-Thm4-zero-rand-cap-1}), we conclude that $S(W)$ (regarded
as a set in $\mathbb{R}^2$) does not intersect the line $L = \left \{ (x, y) \in \mathbb{R}^2:\ x = y \right \}$.  It then follows from the hyperplane separation theorem
\cite{dunford-schwartz-57}
that for some $\eta > 0$, it either holds that
\begin{equation}
S(W) \subseteq \left \{ \left( \alpha, \tilde{\alpha} \right) \in [0, 1]^2:\
\tilde{\alpha} \leq \alpha - \eta \right \},
\end{equation}
or
\begin{equation}
S(W) \subseteq \left \{ \left( \alpha, \tilde{\alpha} \right) \in [0, 1]^2:\
\tilde{\alpha} \geq \alpha + \eta \right \}.
\end{equation}
If the first case happens, then we get from
(\ref{eqn-Pf-Thm4-region-R2-increment}), i.e., $\left( \alpha_k,
\alpha_{k+1} \right) \in S(W),\ k = 0, \ldots, a-1$, that
$\alpha_a \leq \alpha_{a-1} - \eta \leq \ldots \leq \alpha_0 - a
\eta$ which in turn yields that $a \leq \frac{1}{\eta},$ as
$\alpha_a \geq 0$ and $\alpha_0 \leq 1$. Similarly, if the second
case is true, then (\ref{eqn-Pf-Thm4-region-R2-increment}) also
gives that $a \leq \frac{1}{\eta},$ as $\alpha_a \leq 1$ and
$\alpha_0 \geq 0.$

It is now left to prove that under the conditions in the claim,
there are finitely many $u$ for which there exist $U \left( s
\vert x_2, y_2, \ldots, x_{u+1}, y_{u+1} \right)$ that fulfill
(\ref{sym-1}). For any set of nonnegative integers $i, j, k, l$
such that $i + j + k + l = u,$ we let $\alpha_{i, j}^{k, l} = U
\left( 1 \vert \left( 0, 0 \right)^i, \left( 0, 1 \right)^j,
\left( 1, 0 \right)^k, \left( 1,1 \right)^l \right)$. We then
conclude from (\ref{sym-1}) that for any $i > 0,\ j > 0,\ i + j +
k + l = u,$
\begin{equation}
\left [
\bf{w}_{10}^0,
\bf{w}_{10}^1
\right ]
\left [
\begin{array}{cc}
1 - \alpha_{i, j}^{k, l} \\
\alpha_{i, j}^{k, l}
\end{array}
\right ]
=
\left [
\bf{w}_{00}^0,
\bf{w}_{00}^1
\right ]
\left [
\begin{array}{cc}
1 - \alpha_{i-1, j}^{k+1, l} \\
\alpha_{i-1, j}^{k+1, l}
\end{array}
\right ],
\nonumber
\end{equation}
\begin{equation}
\left [
\bf{w}_{11}^0,
\bf{w}_{11}^1
\right ]
\left [
\begin{array}{cc}
1 - \alpha_{i, j}^{k, l} \\
\alpha_{i, j}^{k, l}
\end{array}
\right ]
=
\left [
\bf{w}_{01}^0,
\bf{w}_{01}^1
\right ]
\left [
\begin{array}{cc}
1 - \alpha_{i, j-1}^{k, l+1} \\
\alpha_{i, j-1}^{k, l+1}
\end{array}
\right ].
\label{eqn-Pf-Thm4-region-R3-increment}
\end{equation}
Similar to (\ref{eqn-Pf-Thm4-region-R2}), the set of all triplets $\left( \alpha, \tilde{\alpha}, \hat{\alpha} \right)
\in [0, 1]^3$ that satisfy
\begin{equation}
\left [
\bf{w}_{10}^0,
\bf{w}_{10}^1
\right ]
\left [
\begin{array}{cc}
1 - \alpha \\
\alpha
\end{array}
\right ]
=
\left [
\bf{w}_{00}^0,
\bf{w}_{00}^1
\right ]
\left [
\begin{array}{cc}
1 - \tilde{\alpha} \\
\tilde{\alpha}
\end{array}
\right ],
\nonumber
\end{equation}
\begin{equation}
\left [
\bf{w}_{11}^0,
\bf{w}_{11}^1
\right ]
\left [
\begin{array}{cc}
1 - \alpha \\
\alpha
\end{array}
\right ]
=
\left [
\bf{w}_{01}^0,
\bf{w}_{01}^1
\right ]
\left [
\begin{array}{cc}
1 - \hat{\alpha} \\
\hat{\alpha}
\end{array}
\right ].
\label{eqn-Pf-Thm4-region-R3}
\end{equation}
is closed and convex.  We denote this set by $R(W) \in [0, 1]^3$.  From
(\ref{eqn-Pf-Thm4-zero-rand-cap-1}),
we conclude that $R(W)$ does not intersect the line $\tilde{L} = \left \{
(x, y, z) \in \mathbb{R}^3:\ x = y = z \right \},$ and the hyperplane separation
theorem yields that there exist $a, b, c \in \mathbb{R}$ with $a + b + c = 0,$
and $\eta \geq 0$ such that $R(W) \subseteq \left \{ (x, y, z):\ ax + by + cz \geq \eta \right \}.$  We then consider all possible cases.

Case 1: One of $a, b$ or $c$ is zero.  The subcases $b=0$ and
$c=0$ can be handled similarly, so without loss of generality we
only consider the subcase of $b=0$.  In this subcase, by virtue of
fact that $a+c = 0$, we get that $\left( \alpha, \tilde{\alpha},
\hat{\alpha} \right)$ satisfying (\ref{eqn-Pf-Thm4-region-R3})
must fulfill $\hat{\alpha} \leq \alpha - \epsilon,$ or
$\hat{\alpha} \geq \alpha + \epsilon,$ for some $\epsilon > 0,$
which from (\ref{eqn-Pf-Thm4-region-R3-increment}) yields
\begin{equation}
\alpha_{0, 0}^{0, u} \leq \alpha_{0, 1}^{0, u-1} \leq \ldots
\leq \alpha_{0, u}^{0, 0} - u \epsilon,
\end{equation}
or
\begin{equation}
\alpha_{0, 0}^{0, u} \geq \alpha_{0, 1}^{0, u-1} \geq \ldots
\geq \alpha_{0, u}^{0, 0} + u \epsilon,
\end{equation}
respectively, thereby giving that $u \leq \frac{1}{\epsilon} <
\infty$.  When $a=0,$ we get that any $\left( \alpha,
\tilde{\alpha}, \hat{\alpha} \right)$ satisfying
(\ref{eqn-Pf-Thm4-region-R3}) must fulfil $\hat{\alpha} \leq
\tilde{\alpha} - \epsilon,$ or $\hat{\alpha} \geq \tilde{\alpha} +
\epsilon,$ for some $\epsilon > 0,$ which from
(\ref{eqn-Pf-Thm4-region-R3-increment}), yields
\begin{equation}
\alpha_{ \lceil u/2 \rceil, 0}^{0, \lfloor u/2 \rfloor}
\leq \alpha_{ \lceil u/2 \rceil - \lfloor u/2 \rfloor, \lfloor u/2 \rfloor}
^{\lfloor u/2 \rfloor, 0}  - \lfloor u/2 \rfloor \epsilon,
\end{equation}
or
\begin{equation}
\alpha_{ \lceil u/2 \rceil, 0}^{0, \lfloor u/2 \rfloor}
\geq \alpha_{ \lceil u/2 \rceil - \lfloor u/2 \rfloor, \lfloor u/2 \rfloor}
^{\lfloor u/2 \rfloor, 0}  + \lfloor u/2 \rfloor \epsilon,
\end{equation}
respectively,
thereby giving that $u \leq 2 \left( 1 + \frac{1}{\epsilon} \right) < \infty$.

Case 2.  None of $a, b$ or $c$ is zero.  The two subcases of $ba >
0$ and $ca > 0$ (as $a+b+c = 0,$ one of them must be true) can be
handled similarly; we shall just consider the subcase of $ba > 0$.
From $a+b+c = 0$, we get from $ax + by + cz \geq \eta$ and $ba >
0$ that either $z \leq ax + (1-a) y - \epsilon$ or $z \geq ax +
(1-a)y + \epsilon,$ for some $a \in (0, 1)$ and $\epsilon > 0$.
The two cases can be handled similarly so we shall just show the
first case.  Note that
\begin{equation}
\alpha_{k, \lfloor u/2 \rfloor}^{\lceil u/2 \rceil - k, 0} \leq 1,\ k = 0, \ldots,
\lfloor u/2 \rfloor.
\label{Pf-Thm4-Case2}
\end{equation}
We are in the case when $R(W) \subseteq
\left \{(x, y, z) \in \mathbb{R}^3:\ z \leq ax + (1-a)y - \epsilon \right \}$; hence, we get from (\ref{eqn-Pf-Thm4-region-R3-increment}) and
(\ref{Pf-Thm4-Case2}) that
\begin{eqnarray}
\alpha_{k, \lfloor u/2 \rfloor - 1}^{\lceil u/2 \rceil - k, 1} &\leq& 
a \alpha_{k, \lfloor u/2 \rfloor}^{\lceil u/2 \rceil - k, 0} 
+ \left( 1 - a \right) \alpha_{k - 1, \lfloor u/2 \rfloor}^{\lceil u/2 \rceil - k + 1, 0}  - \epsilon \nonumber \\
&\leq&1 - \epsilon,\ \ \ k = 1, \ldots,
\lfloor u/2 \rfloor.
\end{eqnarray}
If we apply this procedure recursively, we get that for every $j =
1, \ldots \lfloor u/2 \rfloor,$
\begin{equation}
\alpha_{k, \lfloor u/2 \rfloor - j}^{\lceil u/2 \rceil - k, j} \leq 1 - j \epsilon,\ k = j, \ldots,
\lfloor u/2 \rfloor,
\end{equation}
which yields $\alpha_{\lfloor u/2 \rfloor, 0}^{\lceil u/2 \rceil -
\lfloor u/2 \rfloor,\lfloor u/2 \rfloor} \leq 1 - \lfloor u/2
\rfloor \epsilon,$ thereby giving that $u \leq 2 \left(
\frac{1}{\epsilon} +1 \right)$.

Lastly, to show that the symmetrizability of a binary AVMAC can be
arbitrarily large, it suffices to show that for any $N > 0,$ we
can assign the various segments $W_{00}, W_{01}, W_{10},$ and
$W_{11}$ on the simplex so that
(\ref{eqn-Pf-Thm4-zero-rand-cap-1}) and
(\ref{eqn-Pf-Thm4-zero-rand-cap-2}) are fulfilled, and that
(\ref{sym-2}) will be fulfilled for some $U \left(s \vert x_2,
\ldots, x_{a+1} \right),$ and some $a > N.$  This can be done
quite easily by letting $W_{00}$ and $W_{10}$ be of the same
(sufficiently small but positive) length and the same orientation
(it holds that $W(0 \vert 0, 0 ,0) < W(0 \vert 0, 0, 1)$ and $W(0
\vert 1, 0 ,0) < W(0 \vert 1, 0, 1),$ or the other way around with
$>$ instead) but slightly misaligned with each other. Furthermore,
we select $W_{01}$ and $W_{11}$ to be translated versions of
$W_{00}$ and $W_{10},$ (with the same offset) respectively, which
are sufficiently far apart from them so that $W_{00} \cap W_{01} =
\emptyset,$ thereby satisfying
(\ref{eqn-Pf-Thm4-zero-rand-cap-2}). A consequence of this
construction will be that the constraint on the second line of
(\ref{eqn-Pf-Thm4-region-R2}) defining $S(W)$ is redundant and
that $S(W)$ is a line parallel to and, say, above $L = \left \{
(x, y) \in \mathbb{R}^2:\ x = y \right \}$ which is sufficiently
close to it.  Then, starting from $\alpha_0 = 0,$ there can be an
arbitrarily large $a > 0$ (as $S(W)$ gets arbitrarily close to
$L$) such that $\left( \alpha_k, \alpha_{k+1} \right) \in S(W),\ k
= 0, \ldots, a-1,$ and that $\alpha_a \leq 1,$ thereby rendering
an arbitrarily large $a$ for the $U \left( s \vert x_2, \ldots,
x_{a+1} \right)$ satisfying (\ref{sym-2}). $\qed$


\section{\tdred{Discussion}}

At present, there is a gap between Theorem 2 and Theorem 3, i.e.,
there exists a range of list sizes for which we cannot determine
$C_L$.  This is caused by the fact that our present definition of
symmetrizability only captures the ``shape'' of an $A$-diagonal
(\ref{sym-1}) and an $A$-rectangle (\ref{sym-2}), while the output
of a list decoder (for a fixed received sequence) can have any
``shape.''  It is not clear how to capture these
complicated shapes in a single-letter manner as in (\ref{sym-1})
and (\ref{sym-2}).  A full characterization of $C_L$ may entail
a multi-letter formula.

\section{Acknowledgement}
The author thanks Arya Mazumdar for his helpful comment leading to
(\ref{eqn:ProofThm3-3}).

\section{Appendices}

\subsection{Proof of Claim 2}

We prove the claim by contradiction. Denote the set of all
permutations of $[{\Omega}+2]$ by $\mathcal{P}_{{\Omega}+2}$.  
If the claim is false, then for any $\nu > 0$ no matter how small, there exists a
collection of ${\Omega}+2$ distributions $U_i$ for which
(\ref{eqn:Lemma_2}) is violated. Since the left side of
(\ref{eqn:Lemma_2}) is preserved when the indices of $(x_1, y_1),
\ldots, (x_{{\Omega}+2}, y_{{\Omega}+2})$ are permuted, it holds for every $\pi
\in \mathcal{P}_{{\Omega}+2}$ and every $(i, j), 1 \leq i < j \leq {\Omega}+2$,
that

\vspace{0.1in}
$\hspace{-0.15in} \nu >$
\begin{eqnarray}
\ d \left (
\begin{array}{ll}
    P(x_i)Q(y_i) \\
    \left( \sum_{s \in \mathcal{S}} W(z|x_i, y_i, s)
    U_{\pi^{-1}(i)}(x^{\pi([{\Omega}+2])}_{i}, y^{\pi([{\Omega}+2])}_{i}, s) \right), \\
    P(x_j)Q(y_j) \\
    \left( \sum_{s \in \mathcal{S}} W(z|x_j, y_j, s)
    U_{\pi^{-1}(j)}(x^{\pi([{\Omega}+2])}_{j}, y^{\pi([{\Omega}+2])}_{j}, s) \right)
\end{array}
\right ), \nonumber
\end{eqnarray}
where \\
$x^{\pi([{\Omega}+2])}_{i} = (x_{\pi(1)}, \ldots,
x_{\pi(\pi^{-1}(i)-1)}, x_{\pi(\pi^{-1}(i)+1)}, \ldots
x_{\pi({\Omega}+2)})$ and similarly for $y^{\pi([{\Omega}+2])}_i$.

Averaging over $\mathcal{P}_{{\Omega}+2}$ and applying Jensen's
inequality, we get for every $(i, j), 1 \leq i < j \leq {\Omega}+2$, that
\begin{equation}
d \left (
\begin{array}{ll}
    P(x_i)Q(y_i) \\
    \left( \sum_{s \in \mathcal{S}} W(z|x_i, y_i, s)
    U(x^{{\Omega}+2}_{i}, y^{{\Omega}+2}_{i}, s) \right), \\
    P(x_j)Q(y_j) \\
    \left( \sum_{s \in \mathcal{S}} W(z|x_j, y_j, s)
    U(x^{{\Omega}+2}_{j}, y^{{\Omega}+2}_{j}, s) \right)
\end{array}
\right ) < \nu, \label{eqn:App_A_2}
\end{equation}
where

\vspace{0.1in}
$\hspace{-0.15in} U(x^{{\Omega}+2}_i, y^{{\Omega}+2}_i, s)$
\begin{eqnarray}
&\triangleq& \frac{1}{({\Omega}+2)!} \sum_{\pi \in \mathcal{P}_{{\Omega}+2}}
U_{\pi^{-1}(i)}(x^{\pi([{\Omega}+2])}_{i},
y^{\pi([{\Omega}+2])}_{i}, s) \nonumber \\
&=& \frac{1}{({\Omega}+2)!} \sum_{l=1}^{{\Omega}+2} \mathop{\sum_{\pi \in
\mathcal{P}_{{\Omega}+2}}}_{\pi^{-1}(i) = l} U_l(x^{\pi([{\Omega}+2])}_{i},
y^{\pi([{\Omega}+2])}_{i}, s) \nonumber \\
&=& \frac{1}{({\Omega}+2)!} \sum_{l=1}^{{\Omega}+2} \sum_{ \bar{\pi} \in
\mathcal{P}_{{\Omega}+1}} U_l(x^{\bar{\pi}([{\Omega}+2] \backslash \{i\})}_{i},
y^{\bar{\pi}([{\Omega}+2] \backslash \{i\})}_{i}, s). \nonumber
\end{eqnarray}
Clearly, $U$ is symmetric in $(x_1, y_1), \ldots (x_{{\Omega}+1},
y_{{\Omega}+1})$.  Consequently, we conclude that for any $\nu > 0$ no
matter how small, there exists a distribution $U$ for which

\vspace{0.1in}
$\hspace{-0.15in} \max_{1 \leq i < j \leq {\Omega}+2}$
\begin{eqnarray}
\ \ d \left (
\begin{array}{ll}
    P(x_i)Q(y_i) \\
    \left( \sum_{s \in \mathcal{S}} W(z|x_i, y_i, s)
    U(x^{{\Omega}+2}_{i}, y^{{\Omega}+2}_{i}, s) \right), \\
    P(x_j)Q(y_j) \\
    \left( \sum_{s \in \mathcal{S}} W(z|x_j, y_j, s)
    U(x^{{\Omega}+2}_{j}, y^{{\Omega}+2}_{j}, s) \right)
\end{array}
\right ) < \nu. \label{eqn:App_A_3}
\end{eqnarray}

The term on the left side of (\ref{eqn:App_A_3}) is a continuous
function $F(U, P, Q)$ defined on the compact set of all
distributions $U$ on $\mathcal{X}^{{\Omega}+1} \times 
\mathcal{Y}^{{\Omega}+1}
\times \mathcal{S}$ which are symmetric in the sense as mentioned
earlier and all distributions $P$ and $Q$ on $\mathcal{X}$ and
$\mathcal{Y}$ satisfying $\min_{x \in \mathcal{X}} P(x) \geq
\alpha$, $\min_{y \in \mathcal{Y}} Q(y) \geq \alpha$,
respectively. Consequently, there exists $(U^*, P^*, Q^*)$ which
attains the minimum of $F$ and, hence, by (\ref{eqn:App_A_3}), we
have that $F(U^*, P^*, Q^*)=0$. In particular, for any $(i, j),\ 1
\leq i < j \leq {\Omega}+2$, we have that
\begin{eqnarray}
    P^*(x_i)Q^*(y_i)
    \left( \sum_{s \in \mathcal{S}} W(z|x_i, y_i, s)
    U^*(x^{{\Omega}+2}_{i}, y^{{\Omega}+2}_{i}, s) \right) = \nonumber \\
    \ \ P^*(x_j)Q^*(y_j)
    \left( \sum_{s \in \mathcal{S}} W(z|x_j, y_j, s)
    U^*(x^{{\Omega}+2}_{j}, y^{{\Omega}+2}_{j}, s) \right). \label{eqn:App_A_4}
\end{eqnarray}
Marginalizing out the $z$ in (\ref{eqn:App_A_4}), we have that for
every $(i, j),\ 1 \leq i < j \leq {\Omega}+2$ and every $(x_1, y_1),
\ldots, (x_{{\Omega}+2}, y_{{\Omega}+2}),$
\begin{eqnarray}
    P^*(x_i)Q^*(y_i)
    U^*(x^{{\Omega}+2}_{i}, y^{{\Omega}+2}_{i}) = \nonumber \\
    \ \ P^*(x_j)Q^*(y_j)
    U^*(x^{{\Omega}+2}_{j}, y^{{\Omega}+2}_{j}). \label{eqn:App_A_5}
\end{eqnarray}
Clearly, (\ref{eqn:App_A_4}) together with the facts that $\min_{x
\in \mathcal{X}} P^*(x) \geq \alpha$, $\min_{y \in \mathcal{Y}}
Q^*(y) \geq \alpha$ render the required contradiction (since $W$
is not ${\Omega}+1$-symmetrizable) if it holds that
\begin{equation}
    U^*(x^{{\Omega}+1}, y^{{\Omega}+1}) =
    {P^*}^{{\Omega}+1}(x^{{\Omega}+1}){Q^*}^{{\Omega}+1}(y^{{\Omega}+1}). \label{eqn:App_A_6}
\end{equation}
We now show that (\ref{eqn:App_A_6}) indeed follows from
(\ref{eqn:App_A_5}) which is done by induction on ${\Omega}+2$. First,
when ${\Omega}+2 = 2$, (\ref{eqn:App_A_5}) gives that for every $(x_1,
y_1), (x_{2}, y_{2})$, it holds that
\[P^*(x_1)Q^*(y_1)U^*(x_2, y_2) = P^*(x_2)Q^*(y_2)U^*(x_1, y_1),\]
which by summing over $(x_2, y_2)$ gives (\ref{eqn:App_A_6}).  For
every $i < {\Omega}+2$ and $j={\Omega}+2$, marginalizing (\ref{eqn:App_A_5}) with respect to $(x_{{\Omega}+2}, y_{{\Omega}+2})$ gives
\begin{eqnarray} U^*(x^{{\Omega}+1},
y^{{\Omega}+1}) = \hspace{1.0in} \nonumber \\ \tilde{U}^*(x^{{\Omega}+1}_i,
y^{{\Omega}+1}_i)P^*(x_i)Q^*(y_i), \label{eqn:App_A_7}
\end{eqnarray}
where $\tilde{U}^*(x^{{\Omega}+1}_i, y^{{\Omega}+1}_i) = 
\sum_{x_{{\Omega}+2}, y_{{\Omega}+2}}
U^*(x^{{\Omega}+2}_i, y^{{\Omega}+2}_i)$.  Consequently, for every $(i, j),\ 1
\leq i < j \leq {\Omega}+1$, marginalizing (\ref{eqn:App_A_5}) with respect to
$(x_{{\Omega}+2}, y_{{\Omega}+2})$ also gives
\begin{eqnarray} P^*(x_i)Q^*(y_i)
\tilde{U}^*(x^{{\Omega}+1}_i, y^{{\Omega}+1}_i) = 
U^*(x^{{\Omega}+1}, y^{{\Omega}+1}) = \nonumber \\
P^*(x_j)Q^*(y_j)\tilde{U}^*(x^{{\Omega}+1}_j, y^{{\Omega}+1}_j).
\label{eqn:App_A_8}
\end{eqnarray}
By the inductive hypothesis, it follows from (\ref{eqn:App_A_8})
that
\begin{equation}
    \tilde{U}^*(x^{\Omega}, y^{\Omega}) =
    {P^*}^{\Omega}(x^{\Omega}){Q^*}^{\Omega}(y^{\Omega}). \nonumber
\end{equation}
which when combined with (\ref{eqn:App_A_7}) gives
(\ref{eqn:App_A_6}). $\qed$

\subsection{Proof of Claim 3}

We prove the claim by contradiction. Denote the set of all pair of
permutations of $[a+1], [b+1]$ by $\mathcal{P}_{a+1, b+1}$.  If
the claim is false, then for any $\nu > 0$ no matter how small,
there exists a collection of \tdred{$(a+1)(b+1) \geq {\Omega}+2$} distributions
$U_{ij}$ for which (\ref{eqn:Lemma_3}) is violated. Since the left
side of (\ref{eqn:Lemma_3}) is preserved when the indices of
$(x_1, \ldots, x_{a+1})$ and those of $(y_1, \ldots, y_{b+1})$ are
permuted by the first and the second permutations in
$\mathcal{P}_{a+1, b+1}$, respectively, it holds for every
$(\sigma, \pi) \in \mathcal{P}_{a+1, b+1}$ and every $(i, j) \neq
(i', j') \in [a+1] \times [b+1]$, that

\vspace{0.10in}
$\hspace{-0.1in} \nu >$
\begin{eqnarray} \ \ \ d \left (
\begin{array}{ll}
    P(x_i)Q(y_j)\\
    \left( \sum_{s \in \mathcal{S}}
    \vspace{0.02in}
    \begin{array}{ll}
    W(z|x_i, y_j, s) \\
    U_{\sigma^{-1}(i) \pi^{-1}(j)}(x^{\sigma([a+1])}_i, y^{\pi([b+1])}_j, s)
    \end{array} \right),\\
    P(x_{i'})Q(y_{j'})\\
    \left( \sum_{s \in \mathcal{S}}
    \vspace{0.02in}
    \begin{array}{ll}
    W(z|x_{i'}, y_{j'}, s) \\
    U_{\sigma^{-1}(i') \pi^{-1}(j')}(x^{\sigma([a+1])}_{i'}, y^{\pi([b+1])}_{j'},
    s)
    \end{array}
    \right)
\end{array}
\right ). \nonumber
\end{eqnarray}
\vspace{0.05in}
where \\
$x^{\sigma([a+1])}_{i} = (x_{\sigma(1)},
\ldots, x_{\sigma(\sigma^{-1}(i)-1)},
x_{\sigma(\sigma^{-1}(i)+1)}, \ldots x_{\sigma(a+1)})$ and
similarly for $y^{\pi([b+1])}_i$.

Averaging over $\mathcal{P}_{a+1, b+1}$ and applying Jensen's
inequality, we get for every $(i, j) \neq (i', j') \in [a+1]
\times [b+1]$, that
\begin{eqnarray} d \left (
\begin{array}{ll}
    P(x_i)Q(y_j)\\
    \left( \sum_{s \in \mathcal{S}}
        W(z|x_i, y_j, s)
        U(x^{a+1}_i, y^{b+1}_j, s) \right),\\
    P(x_{i'})Q(y_{j'})\\
    \left( \sum_{s \in \mathcal{S}}
        W(z|x_{i'}, y_{j'}, s)
        U(x^{a+1}_{i'}, y^{b+1}_{j'}, s) \right)
    \end{array}
\right ) \ <\ \nu. \label{eqn:App_B_2}
\end{eqnarray}
where

$\hspace{-0.1in} (a+1)! (b+1)! U(x^{a+1}_i, y^{b+1}_j, s)$
\begin{eqnarray}
&\triangleq& \sum_{\sigma, \pi \in \mathcal{P}_{a+1, b+1}}
U_{\sigma^{-1}(i) \pi^{-1}(j)}(x^{\sigma([a+1])}_i,
y^{\pi([b+1])}_j, s)
\nonumber \\
&=& \sum_{u=1}^{a+1} \sum_{v=1}^{b+1} \mathop{\sum_{(\sigma, \pi)
\,\in\, \mathcal{P}_{a+1, b+1}}}_{\sigma^{-1}(i) = u,\ \pi^{-1}(j)
= v} U_{u, v}(x^{\sigma([a+1])}_{i},
y^{\pi([b+1])}_{j}, s) \nonumber \\
&=& \sum_{u=1}^{a+1} \sum_{v=1}^{b+1} \sum_{(\bar{\sigma},
\bar{\pi}) \in \mathcal{P}_{a, b}} U_{u, v}(x^{\bar{\sigma}([a+1]
\backslash \{i\})}_{i}, y^{\bar{\pi}([b+1] \backslash \{j\})}_{j},
s). \nonumber
\end{eqnarray}
Clearly, $U$ is symmetric in $(x_1, \ldots, x_{a})$ and $ (y_1,
\ldots, y_{b})$. Consequently, we conclude that for any $\nu > 0$
no matter how small, there exists a distribution $U$ for which

\vspace{0.15in}
$\hspace{-0.15in} \max_{(i, j) \neq (i', j') \in [a+1] \times
[b+1]}$
\begin{eqnarray}
\ \ d \left (
\begin{array}{ll}
    P(x_i)Q(y_j)\\
    \left( \sum_{s \in \mathcal{S}}
        W(z|x_i, y_j, s)
        U(x^{a+1}_i, y^{b+1}_j, s) \right),\\
    P(x_{i'})Q(y_{j'})\\
    \left( \sum_{s \in \mathcal{S}}
        W(z|x_{i'}, y_{j'}, s)
        U(x^{a+1}_{i'}, y^{b+1}_{j'}, s) \right)
\end{array}
\right ) < \nu. \label{eqn:App_B_3}
\end{eqnarray}

The term on the left side of (\ref{eqn:App_B_3}) is a continuous
function of $F(U, P, Q)$ defined on the compact set of all
distributions $U$ on $\mathcal{X}^{a} \times \mathcal{Y}^{b}
\times \mathcal{S}$ which are symmetric in the sense as mentioned
earlier and all distributions $P$ and $Q$ on $\mathcal{X}$ and
$\mathcal{Y}$ satisfying $\min_{x \in \mathcal{X}} P(x) \geq
\alpha$, $\min_{y \in \mathcal{Y}} Q(y) \geq \alpha$,
respectively. There exists $(U^*, P^*, Q^*)$ which attains the
minimum of $F$ and, hence, by (\ref{eqn:App_B_3}), we have that
$F(U^*, P^*, Q^*)=0$. In particular, for any $(i, j) \neq (i', j')
\in [a+1] \times [b+1]$, we have that
\begin{eqnarray}
    P^*(x_i)Q^*(y_j)
    (\sum_{s \in \mathcal{S}} W(z|x_i, y_j, s)
    U^*(x^{a+1}_{i}, y^{b+1}_{j}, s)) = \nonumber \\
    \ \ P^*(x_{i'})Q^*(y_{j'})
    (\sum_{s \in \mathcal{S}} W(z|x_{i'}, y_{j'}, s)
    U^*(x^{a+1}_{i'}, y^{b+1}_{j'}, s)). \label{eqn:App_B_4}
\end{eqnarray}

Marginalizing out the $z$ in (\ref{eqn:App_B_3}), we have that for
every $(i, j) \neq (i', j') \in [a+1] \times [b+1]$ and every
$(x_1, \ldots, x_{a+1})$ and $(y_1, \ldots, y_{b+1})$
\begin{eqnarray}
    P^*(x_i)Q^*(y_j)
    U^*(x^{a+1}_{i}, y^{b+1}_{j}) = \nonumber \\
    \ \ P^*(x_{i'})Q^*(y_{j'})
    U^*(x^{a+1}_{i'}, y^{b+1}_{j'}). \label{eqn:App_B_5}
\end{eqnarray}
Clearly, (\ref{eqn:App_B_4}) together with the facts that $\min_{x
\in \mathcal{X}} P^*(x) \geq \alpha$, $\min_{y \in \mathcal{Y}}
Q^*(y) \geq \alpha$ render the required contradiction (since $W$
is not $\Omega+1$-symmetrizable and $(a+1)(b+1) \geq \Omega+2$) if it holds
that for every $(i, j) \neq (i', j') \in [a+1] \times [b+1]$,
\begin{eqnarray}
    P^*(x_i)Q^*(y_j)
    U^*(x^{a+1}_{i}, y^{b+1}_{j}) = \nonumber \\
    {P^*}^{a+1}(x^{a+1}){Q^*}^{b+1}(y^{b+1}). \label{eqn:App_B_6}
\end{eqnarray}

We now show that (\ref{eqn:App_B_6}) indeed follows from
(\ref{eqn:App_B_5}) by induction on $a + b$. First, when $a+b =
1$, (\ref{eqn:App_B_5}) gives that, depending on whether $a = 1,\
b = 0$ or $a = 0,\ b = 1$,
\[P^*(x_1)Q^*(y)U^*(x_2) = P^*(x_2)Q^*(y) U^*(x_1),\ \ \mbox{or}\]
\[P^*(x)Q^*(y_1)U^*(y_2) = P^*(x)Q^*(y_2) U^*(y_1).\]
from which, by $\min_{x \in \mathcal{X}} P^*(x) \geq \alpha$,
$\min_{y \in \mathcal{Y}} Q^*(y) \geq \alpha$, (\ref{eqn:App_B_6})
follows.

Next, without loss of generality, we can assume that $a > 0$. For
$i \in [a]$, substituting $i'=a+1$ and $j'=j$ in
(\ref{eqn:App_B_5}) and marginalizing with respect to $x_{a+1}$
therein together with the fact that $Q^*(y_j) > 0$, we get that
\begin{equation}
    \tilde{U}^*(x_i^a, y_j^{b+1}) P^*(x_i) = U^*(x^a, y_j^{b+1}),
    \label{eqn:App_B_7}
\end{equation}
where $\tilde{U}^*(x_i^a, y_j^{b+1}) = \sum\limits_{x_{a+1}} U^*
\left( x^{a+1}_i, y^{b+1}_{j} \right).$ Consequently, we get from
(\ref{eqn:App_B_7}) that for every $(i, j) \neq (i', j') \in [a]
\times [b+1]$,
\begin{equation}
    \tilde{U}^*(x_i^a, y_j^{b+1}) P^*(x_i) = U^*(x^a, y_j^{b+1}),\ \mbox{and}   \label{eqn:App_B_8}
\end{equation}
\begin{equation}
    \tilde{U}^*(x_{i'}^a, y_{j'}^{b+1}) P^*(x_{i'}) = U^*(x^a, y_{j'}^{b+1}).   \label{eqn:App_B_9}
\end{equation}
Hence,
\begin{equation}
    \tilde{U}^*(x_i^a, y_j^{b+1}) P^*(x_i) Q^*(y_j)= U^*(x^a, y_j^{b+1}) Q^*(y_j),\ \mbox{and}   \label{eqn:App_B_10}
\end{equation}
\begin{equation}
    \tilde{U}^*(x_{i'}^a, y_{j'}^{b+1}) P^*(x_{i'}) Q^*(y_{j'})= U^*(x^a, y_{j'}^{b+1}) Q^*(y_{j'}).
\label{eqn:App_B_11}
\end{equation}
By letting $i=i'=a+1$ in (\ref{eqn:App_B_5}) and using the fact that
$P^*(x_{a+1}) > 0$, we get that for every
$(i, j) \neq (i', j') \in [a] \times [b+1]$,

\vspace{0.1in}
$\hspace{-0.15in} \tilde{U}^*(x_i^a, y_j^{b+1}) P^*(x_i) Q^*(y_j) $
\begin{eqnarray}
    &=& U^*(x^a, y_j^{b+1}) Q^*(y_j) \nonumber \\
    &=& U^*(x^a, y_{j'}^{b+1}) Q^*(y_{j'}),\ \ \mbox{by~}(\ref{eqn:App_B_5}) \nonumber \\
    &=&     \tilde{U}^*(x_{i'}^a, y_{j'}^{b+1}) P^*(x_{i'}) Q^*(y_{j'}).  \label{eqn:App_B_12}
\end{eqnarray}
By the inductive hypothesis, it follows from (\ref{eqn:App_B_12}) that
\[\tilde{U}^*(x^{a-1}, y^b) = {P^*}^{a-1}(x^{a-1}) {Q^*}b(y^b)\]
which when combined with (\ref{eqn:App_B_7}) gives (\ref{eqn:App_B_6})  $\qed$

\subsection{Proof of Lemma 2}

The proof here is based on the proof of Lemma 2 of
\cite{Ahlswede-Cai_99}.  In fact, we use directly the following
proposition from \cite{Ahlswede-Cai_99} and omit its proof.

\vspace{0.05in} \textbf{Proposition:} \cite{Ahlswede-Cai_99}
{\em
For rvs $A_0, A_1, \ldots, A_m$ and functions $f_i(A_0, \ldots,
A_i)$ satisfying $0 \leq f_i \leq 1,\ 1 \leq i \leq m$, if for
every $i = 1, \ldots, m$,
\begin{equation}
E[f_i(A_0, \ldots, A_i) | A_0, \ldots, A_{i-1}] \leq a,\
\mbox{a.s.}, \label{eqn:ProofLemma2-1}
\end{equation}
then for $b > 0$, it holds that
\begin{equation}
Pr\{\sum_{i = 1}^m f_i(A_0, \ldots, A_i)
> mb \} \leq \left(\frac{e}{2} \right)^a 2^{-m(b - a \log_2{e})}.
\nonumber
\end{equation}
}
\vspace{0.05in}

For arbitrary sequences ${\bf x} \in \mathcal{X}^n$ and ${\bf y}
\in \mathcal{Y}^n$ with types $P_{({\bf x})} = P_X$ and $P_{({\bf
y})} = P_Y$, respectively, let ${\bf U}_i, {\bf V}_j,\ i = 1,
\ldots, M,\ j = 1 \ldots, M,$ be independent and uniformly
distributed rvs taking values in $\mathcal{T}_X$ and
$\mathcal{T}_Y$ respectively.

Then, it follows exactly as (18) of \cite{Ahlswede-Cai_99} that
the probability that (\ref{eqn:ProofThm3-12}) is violated (for
some ${\bf s}$) is going to zero. We now prove that the
probability that (\ref{eqn:ProofThm3-13}) is violated (for some
${\bf s}$) is also going to zero. By the symmetry of
$\mathcal{B}_{\epsilon}({\bf s})$ and $\mathcal{C}_{\epsilon}({\bf
s})$ and by the fact that $\vert \mathcal{S}^n \vert$ grows
exponentially with $n$, it suffices to prove that for every ${\bf
s} \in \mathcal{S}^n$, the probability of the event that
\begin{equation}
    \mathcal{B}^c_{\epsilon}({\bf s}) \geq 2^{-\frac{\epsilon}{4} n} M.
    \label{eqn:ProofLemma2-2}
\end{equation}
goes to zero doubly exponentially. To this end, for any collection
of rvs $(X, X_1, \ldots, X_L, Y_1, \ldots, Y_{L+1}, S)$ on
$\mathcal{X}^{L+1} \times \mathcal{Y}^{L+1} \times \mathcal{S}$
with joint distribution being a joint type of some tuples ${\bf x}'
\in \mathcal{X}^n,\ {\bf x}'_i \in \mathcal{X}^n,\ i = 1, \ldots,
L,$ (with the type of each of them being $P_X$),
$\ {\bf y}'_j \in \mathcal{Y}^n,\ j = 1, \ldots L + 1,$
(with the type of each of them being $P_Y$),
$\ {\bf s}'
\in \mathcal{S}^n$, satisfying
\begin{equation}
I(X \wedge X^L, Y^{L+1}, S) > (2L+1)R + \epsilon,
\label{Pf-Lemma2-a}
\end{equation}
and any ${\bf s} \in \mathcal{S}^n$, let
\begin{equation}
    f_i({\bf v}^M, {\bf u}_1, \ldots, {\bf u}_i) =
    \left \{
        \begin{array}{ll}
            1,  & \begin{array}{ll}
                        \mbox{if~} \exists I \subset [i-1],\ |I| = L
                        \mbox{~and}\\
                        J \subset [M],\ |J| = L + 1
                        \mbox{~with~}\\
                        ({\bf u}_i, {\bf u}_I, {\bf v}_J, {\bf s}) \in
                        \mathcal{T}_{X, X^L, Y^{L+1}, S}
                  \end{array}\\
            0   & \mbox{otherwise.}
        \end{array}
    \right. \nonumber
\end{equation}
Using the substitution $A_0 = {\bf V}^M,\ A_i = {\bf U}_i,\ m =
M,\ a = M^{2L+1} 2^{-n(I(X \wedge X^L, Y^{L+1}, S) -
\frac{\epsilon}{4})}$ and $b = 2^{-\frac{\epsilon}{4} n}$ in the
proposition, it follows, using (\ref{eqn:types-3}), that
(\ref{eqn:ProofLemma2-1}) holds, i.e.,
\[\hspace{-1.6in} E[f_i(A_0, \ldots, A_i) | A_0, \ldots, A_{i-1}]\]
\begin{eqnarray}
&\leq& {\tiny \left(
\begin{array}{cc} i - 1 \\ L \end{array} \right)
\left(
\begin{array}{cc} M \\ L + 1 \end{array} \right)}
2^{-n D(P_{X, X^L, Y^{L+1}, S}|| P_{X} \times P_{X^L, Y^{L+1},
S})} \nonumber \\
&\leq& M^{2L+1}
2^{-n I(X \wedge X^L, Y^{L+1}, S)} \nonumber \\
&\leq& a
\ \ \leq\ \ 2^{-\frac{3 \epsilon}{4} n},\ \mbox{a.s.},
\ \ \mbox{by~}
(\ref{Pf-Lemma2-a}). \nonumber
\end{eqnarray}
Since there are only polynomially many joint types satisfying
(\ref{Pf-Lemma2-a}), we get the doubly exponential decay of the
probability of the event in (\ref{eqn:ProofLemma2-2}) from the
proposition. $\sqr$

\newpage
Sirin Nitinawarat obtained the B.S.E.E. degree from Chulalongkorn
University, Bangkok, Thailand, with first class honors,
and the M.S.E.E. degree from the University of Wisconsin, Madison. 
He received his Ph.D. degree from the Department of Electrical and Computer
Engineering and the Institute for Systems Research at the
University of Maryland, College Park, in December 2010.  He is now a posdoctoral 
research associate at the University of Illinois at Urbana-Champaign and the Coordinated 
Science Laboratory.  His research interests are in information and coding theory, communications, statistical signal processing, estimation and detection, stochastic control, and machine learning.  

Dr. Nitinawarat was Co-Organizer for the special session on {\em Controlled Sensing for 
Inference} at the 2012 IEEE International Conference on Acoustics, Speech and Signal Processing (ICASSP); chair for the session on {\em Distributed Inference in Sensor Networks} at the $49^{th}$ Annual Allerton Conference on Communication, Control, and Computing (2011).  He was a finalist for the best student paper award for the 
IEEE International Symposium on Information Theory which was held at Austin, 
Texas in 2010.  


\begin{thebibliography}{99}

\bibitem{Ahlswede_78} R. Ahlswede, ``Elimination of correlation in
random codes for arbitrarily varying channels,'' {\em Z.
Wahrscheinlichkeitsrechnung verw. Geb.}, vol.~44, pp.~159--175,
1978.


\bibitem{Ahlswede-Cai_99} R. Ahlswede and N. Cai, ``Arbitrarily
varying multiple-access channels, part I--Ericson's
symmetrizability is adequate, Gubner's conjecture is true,'' {\em
IEEE Trans. Inf. Theory,} vol.~45, no.~2, pp.~742--749, Mar.~1999.

\bibitem{Blinovsky-Narayan-Pinsker_95} V. Blinovsky, P. Narayan and
M. Pinsker, ``Capacity of the arbitrarily varying channel under
list decoding,'' {\em Probl. Pered. Inform.,} vol.~31, no.~2, pp.
99--113, 1995.

\bibitem{Csiszar_Korner_81} I. Csisz\'ar and J. K\"orner, {\it
Information Theory: Coding Theorems for Discrete Memoryless
Systems,} Akad\'emiai Kiad\'o, Budapest~1981.

\bibitem{Csiszar_Narayan_88} I. Csisz\'ar and P. Narayan, `` The
capacity of the arbitrarily varying channel revisited: positivity,
constraints,'' {\em IEEE Trans. Inf. Theory,} vol.~34,
pp.~181--193, Mar.~1988.

\bibitem{dunford-schwartz-57} N. Dunford and J. T. Schwartz, {\em
Linear Operators: Part I,} Interscience, New York, 1951.

\bibitem{Ericson_85} T. Ericson, ``Exponential error bounds for random
codes in arbitrarily varying channels,'' {\em IEEE Trans. Inf.
Theory,} vol.~31, pp.~42--48, Jan.~1985.

\bibitem{Gubner_90} J. A. Gubner, ``On the deterministic-code capacity
of the multiple-access arbitrarily varying channel,'' {\em IEEE
Trans. Inf. Theory,} vol.~36, pp.~262--275, Mar.~1990.

\bibitem{Hughes_97} B. L. Hughes, ``The smallest list for the arbitrarily
varying channel,'' {\em IEEE Trans. Inf. Theory,} vol.~43,
pp.~803--815, May~1997.

\bibitem{Jahn_81} J-H. Jahn, ``Coding of arbitrarily varying multiuser
channels,'' {\em IEEE Trans. Inf. Theory,} vol.~27, pp.~212--226,
May~1981.

\end{thebibliography}
\end{document}